# Accepted manuscript

# Influence of on-site low-ureolysis bacteria and high-ureolysis bacteria on the effectiveness of MICP processes

Qinghua Wu[a], Yuze Wang[b]*

a.  Research Assistant; Department of Ocean Science and Engineering, Southern University of Science and Technology, Shenzhen, 518055, China; Email: wuqh@mail.sustech.edu.cn

b.  Associate Professor; Department of Ocean Science and Engineering, Southern University of Science and Technology, Shenzhen, 518055, China; Email: wangyz@sustech.edu.cn

*Corresponding author: wangyz@sustech.edu.cn

**Abstract：**

Microbially Induced Calcium Carbonate Precipitation (MICP) is an emerging eco-friendly microbial engineering technique that utilizes urease-producing microorganisms to enhance the mechanical properties of soils. *Sporosarcina pasteurii* (*S. pasteurii*) stands out among these microorganisms as an efficient urease producer. However, field trials often lead to less-than-optimal experimental outcomes due to the effects of presence of native soil microbes. To evaluate the impact of indigenous microorganisms on the effectiveness of MICP at the site, bacteria isolated from natural soil, classified into groups of on-site low-ureolysis and high-ureolysis bacteria (OSLUB and OSHUB, respectively), were combined with *S. pasteurii* to conduct MICP experiments both in microfluidic chips and sand columns. Analysis covered bacterial population, urease activity, pH changes, calcium carbonate crystal count and volume, as well as the unconfined compressive strength (UCS) of reinforced samples. Experimental results revealed that combining OSLUB (on-site low-ureolysis bacteria)with *S. pasteurii* led to a reduction in bacterial activity of 74-84% by 120 hours, resulted in an approximately 60% decrease in chemical conversion rate and the UCS of MICP-treated soils was 60% lower than the *S. pasteurii*. However, when OSHUB (on-site high-ureolysis bacteria) is mixed with S. pasteurii, although there is a reduction in bacterial activity by 49-54% by the 120-hour mark, it remains less pronounced than the activity decrease observed in *S. pasteurii* alone, which is 64%. Consequently, the rates of calcium carbonate chemical conversion were enhanced by 9% to 45%, and the UCS of the reinforced sand columns showed a slight improvement relative to the control group. This research highlights the distinct impacts of OSLUB and OSHUB on the efficiency of MICP on location. The main difference between OSLUB and OSHUB lies in their respective effects on pH levels following mixing. OSLUB tends to decrease the pH level gradually in the combined bacterial environment, while OSHUB, in contrast, increases the pH level over time in the same setting. The maintenance of both high bacterial activity and high precipitation rates is crucially dependent on pH levels, highlighting the importance of these findings for enhancing MICP efficiency in field applications. Strategies that either diminish the presence of OSLUB while augmenting that of OSHUB, or that sustain a relatively high pH level, could be valuable. These avenues promise significant improvements and merit further investigation in future studies.

# 1. Introduction

Microbially Induced Calcium Carbonate Precipitation (MICP) as an emerging eco-friendly biogeotechnique has attracted increasing research attention in the past two decades. Throughout the extensive exploration of MICP across multiple scales, a significant portion of research has centered around the urea hydrolysis process (Martinez and DeJong, 2009; Bang et al., 2011; Van Paassen, 2011; Rahman et al., 2020; Chen et al., 2022), mainly due to its rapid reaction and simplified operation. Within this process, microorganisms capable of generating urease enzymes catalyze the hydrolysis of urea, such as S. pasteurii, are introduced into the soil. Subsequently, a cementation solution containing urea and calcium chloride is injected, leading to in-situ calcium carbonate formation (Equations 1 and 2). This process is called bioaugmentation MICP.

$$CO(NH_2)_2 + 2H_2O \xrightarrow{urease} 2NH_4^+ + CO_3^{2-} \qquad (1)$$

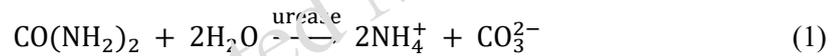

$$Ca^{2+} + CO_3^{2-} \rightarrow CaCO_3 \qquad (2)$$

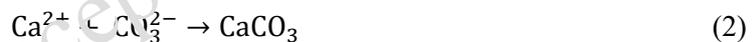

The formed $CaCO_3$ crystals serve as bridging agents between soil particles, thereby cementing the natural soil particles together; On the other hand, the crystals do not entirely occupy the soil pores, thus maintaining a certain level of permeability in the soil structure (DeJong et al., 2006; Whiffin et al., 2007; Qabany and Soga, 2014; Montoya and DeJong, 2015). Therefore, MICP can effectively enhance the shear strength, stability, and liquefaction resistance of the soil structure. The natural environment harbors a diverse array of microbial communities, encompassing bacteria, archaea, algae, fungi, and other prokaryotic and eukaryotic organisms. Microbial populations in the surface soil are approximately $10^6 - 10^{11}$ kg$^{-1}$, while at depths of 2 - 30 meters, microbial populations are around $10^9 - 10^{10}$ kg$^{-1}$ (Mitchell and Santamarina, 2005; Martinez et al., 2013). These environmental microorganisms can exert intricate influences on the bioaugmentation MICP process. Gat et al. (2014) discovered that when other bacteria such as *Bacillus subtilis* was co-cultured with *S. pasteurii*, the growth rate of *Bacillus subtilis* was significantly higher than that of *S. pasteurii*, resulting in greater consumption of nutrients (Gat et al., 2014). Gomez et al. (2015) observed in field-scale MICP experiments that artificially introduced *S. pasteurii* were influenced and inhibited by on-site bacteria at the experimental site. The population of *S. pasteurii* rapidly declined to zero, leading to significantly poorer cementation effects compared to laboratory experiments conducted under the same conditions (Gomez et al., 2015). Murugan et al. (2021) studied the impact of the naturally urea-degrading microbial community (NUMC) on the potential of *S. pasteurii* for MICP. The kinetics of calcium carbonate precipitation for *S. pasteurii* mixed with an eightfold concentration of NUMC exhibited similarity to that without the incorporation of NUMC. However, the rate of transformation of calcium carbonate crystals from aragonite to calcite decreased.

To address the limitations of bioaugmentation MICP mentioned earlier, researchers have proposed a method known as biostimulation MICP (Fujita et al., 2000; Burbank et al., 2011; Gomez et al., 2015; Gomez et al., 2018; Chen and Achal, 2019; Sayed et al., 2021). In this approach, nutrients are injected into the soil to stimulate the growth of on-site urease-producing bacteria. Subsequently, a cementation solution is introduced to produce calcium carbonate. The advantage of biostimulation MICP is that it utilizes local microorganisms, eliminating the concern of non-adaptation to the environment. However, this technique also faces challenges. Firstly, as biostimulation involves the requirement of ensuring the presence of microorganisms with urease characteristics within the experimental site, biostimulation MICP places considerable demands on the suitability of the experimental location, which may not be attainable for all sites. Secondly, current research indicates that the efficiency of inducing calcium carbonate by locally stimulated bacterial growth remains lower compared to artificially cultured strains in laboratory settings (Hammes et al., 2003; Burbank et al., 2013; Zhu and Dittrich, 2016). Research into the effects of native bacteria on the efficiency and biostimulation of MICP emphasizes the necessity of accounting for local microbial populations in developing MICP methods. While prior studies have been initiated, a detailed understanding of the specific local bacterial species that lead to a swift reduction in the effectiveness of introduced *S. pasteurii*, and their influence on the mineralization outcomes of *S. pasteurii*-led MICP, remains elusive. Hence, this investigation aims to bridge this gap by executing an extensive set of experiments to assess the impact of on-site low-ureolysis and high-ureolysis bacteria on both the functionality of *S. pasteurii*-mixed bacterial systems and the overall MICP mineralization efficiency. The insights gained will inform the creation of enhanced MICP strategies that incorporate the dynamics of native bacteria, ultimately improving MICP's application effectiveness in actual soil environments under natural conditions.

## 2. Materials and Methods

**Isolation and identification of on-site bacteria**

Adopting the methodology outlined by Liu et al. (2021) for isolating and identifying bacteria from natural soil, we began with a 10-gram soil sample which was serial diluted in sterile tubes using 90 ml of deionized water, creating dilutions from $10^{-1}$ to $10^{-5}$. These dilutions were then spread onto $NH_4$-YE agar plates prepared with artificial seawater ingredients (Yeast extract, ammonium sulfate, Tris base, and agar) and incubated at 30°C for a duration of 48 hours. This process resulted in the formation of colonies on the plates, each of which was transferred to new $NH_4$-YE agar plates for further growth, totaling 40 isolated colonies. To identify urease-positive bacteria among these isolates, the colonies were next streaked onto a urease-specific agar medium containing peptone, glucose, sodium chloride, potassium dihydrogen phosphate, urea, phenol red, and agar, followed by incubation at 35°C for 2 hours. 30-35°C was selected because bacterial growth rate in these temperature range was suggested to be highest (Vaskevicius et al. 2023). The presence of urease-producing bacteria was indicated by a change in the medium's pH due to urea breakdown into ammonium ions, observable through a color transition from yellow to purple as shown in Figure 1a. Colonies that did not induce a color change were categorized as low urease producers. In parallel, the urease activity of the 40 isolated bacterial strains was quantified as depicted in Figure 1b, using the conductivity method given by Whiffin (2004), in which the reduction of urea by ureolysis and producing of ammonium and carbonate (Equitation 1) was measured based on the conductivity change of the solution. In this procedure, 1.11 M urea solution and the bacterial suspension was mixed at a 9:1 ratio for 5 minutes at 25°C. Electrical conductivity of the mixed solution was measured right after mixing and 5 minutes post-mixing, and the ureolysis rate is calculated using the Equation (3):

$$BA(mM/h) = \frac{\Delta Conductivity\,(\mu S/cm)}{\Delta t\,(\min)} \times \left(10^{-3} \times 11.11\right)\left(mM/(\mu S/cm)\right) \times 60\,(\min/h)$$

(3)

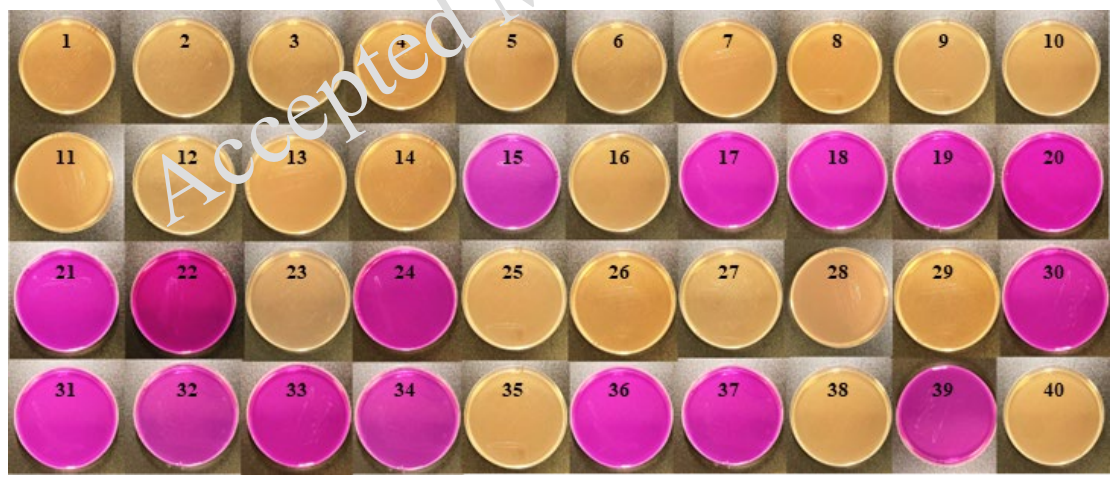

(a)

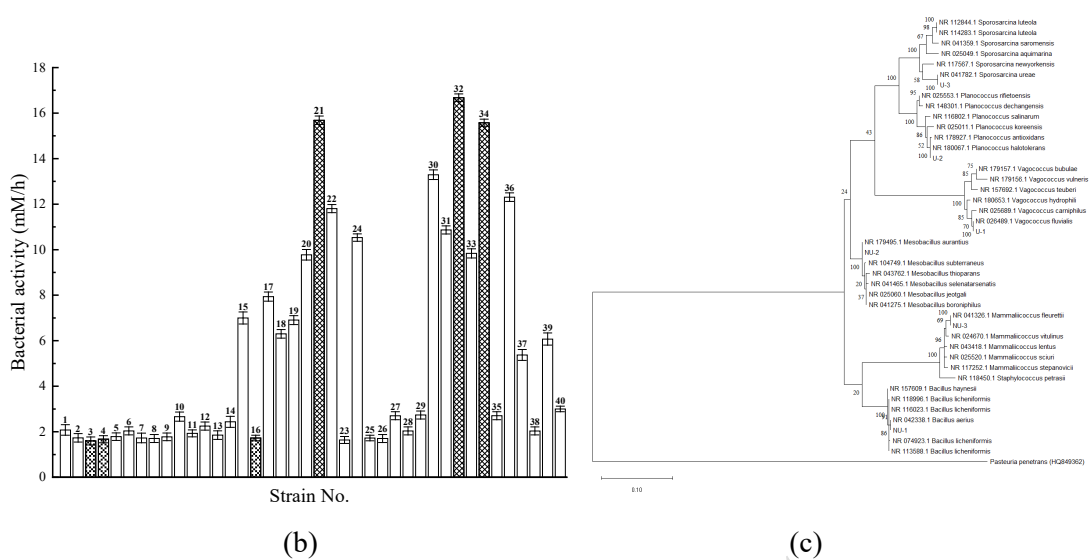

(b)  (c)

Fig. 1. Isolation and purification of bacteria from in-situ soil samples: (a) images from the urease detection experiment (OSHUB appearing purple, OSLUB appearing yellow); (b) results of in-situ bacterial activity testing; (c) phylogenetic trees for the six target bacterial strains.

Notably, the reaction of low-urease bacteria with the urea solution could also yield minimal conductivity changes due to the urea solution's inherent ionization and decomposition. Based on Figure 1b three strains with the highest urease activity, namely high-ureolysis bacteria, and three with the lowest urease activity, namely low-ureolysis bacteria (as shown in Table 1) were selected for further analysis. These selected strains underwent 16S rDNA sequencing to characterize and identify them. The sequence data obtained were then cross-referenced with the GenBank and Apollon DB-BA 9.0 databases for accurate species identification, as shown in Figure 1c. It should be noted that *Bacillus subtilis* is defined as low-ureolysis bacteria in the current study. Whether *Bacillus subtilis* possesses high urease activity and can be effectively utilized in MICP is controversial according to existing research. In the study by Yamasamit et al. (2023), *Bacillus subtilis* is reported to exhibit urease activity and can generate calcium carbonate. However, in other studies by Saricicek et al. (2019) and Chen et al. (2021), *Bacillus subtilis* is regarded as a low-ureolysis bacterium. Hoffmann et al. (2021) investigates demonstrated that although *Bacillus subtilis* possesses the core structural genes ureABC for urease production, it lacks any auxiliary genes, which results in extremely low inherent urease activity in this strain, and without appropriate genetic modification, this strain is unlikely to effectively induce calcium carbonate precipitation. Following the previous research and along with the ureolysis activity tested herein, in this study, *Bacillus subtilis* was considered as a low-ureolysis bacteria.

Table 1 Microbial species in microfluidic chip experiments

| Bacterial species | Urease activity |
| --- | --- |
| *Bacillus subtilis* | OSLUB |
| *Mesobacillus aurantius* | OSLUB |
| *Mammaliicoccus fleurettii* | OSLUB |
| *Vagococcus fluvialis* | OSHUB |
| *Planococcus halotolerans* | OSHUB |
| *Sporosarcina ureae* | OSHUB |
| *Sporosarcina pasteurii* | HUB |

**Bacterial culture and cementation solution**

*S. pasteurii*, as the most commonly used bacterial strain for MICP, was applied in the current study. The *S. pasteurii* strain utilized in this study was purchased in freeze-dried stocks from CGMCC (CGMCC1.3687). The defrosting and cultivation procedures of *S. pasteurii* were conducted following the method presented by Wang et al. (2019b). Because the bacterial growth rate of *S. pasteurii* is highest at 30°C (Chen et al. 2022), the cultivation of *S. pasteurii* was conducted at 30°C. Enrichment of the six target on-site bacteria was carried out in sterile NH$_4$-YE medium, composed of yeast extract (20 g/L), ammonium sulfate (10 g/L), and Tris base (15.74 g/L). Under aerobic conditions, the cultures were incubated at 30°C, following the temperature of *S. pasteurii*, with a shaking speed of 180 rpm until the bacterial culture reached to an optical density measured at a wavelength of 600 nm (OD$_{600}$) of between 1.8-2.5. After cultivations, all bacterial strains were stored at 4°C before the liquid experiment, microfluidic chip experiment or soil column experiment. For these experiments, the bacterial suspension was diluted to diluted by using NH$_4$-YE medium to 1.8±0.2.

The composition of the cementation solution used for MICP treatment in the microfluidic chip experiment and soil column experiment includes calcium chloride, urea, and nutrient broth. Urea and calcium chloride serve as the source materials for forming calcium carbonate precipitates, while nutrient broth provides the energy source for bacterial urease production. The concentrations of calcium chloride, urea, and nutrient broth are 0.5 mM, 0.75 mM, and 3 g/L, respectively. All chemicals used are of analytical grade purity.

**Bacterial mixing liquid experiment**

To study the effects of on-site low-ureolysis bacteria and high-ureolysis bacteria on the bacterial density and activity on the mixed bacterial suspension including both on-site bacteria and *S. pasteurii*, a liquid experiment was conducted. In this experiment, the enriched cultures of the six target on-site bacteria and *S. pasteurii* were all firstly diluted by using NH$_4$-YE medium to 1.8±0.2, and then the six target on-site bacteria were mixed with *S. pasteurii* culture in a 1:1 volume ratio. Each experimental group consisted of three parallel samples, with a total volume of 100 ml per sample. A separate group with unmixed *S. pasteurii* culture was set as the control. At designated time intervals (2 h, 4 h, 6 h, 12 h, 24 h, 48 h, 72 h, 96 h, and 120 h), the bacterial optical density (OD$_{600}$) and bacterial activity (BA) of each mixed bacterial culture were measured and recorded (Wang et al., 2023). During the liquid experiment procedures, all experimental and control samples were placed at room temperature (20°C).

When analyzing bacterial activity data, since the bacterial cultures were mixed in a 1:1 volume ratio, the bacterial counts of each species per unit volume were reduced to half of the original control culture. The control group consisted of *S. pasteurii*, and for comparative analysis, the control group's BA data needed to be halved before comparison with the experimental groups.

**Microfluidic chip experiment, imaging and image analysis**

The microfluidic chip used in this study follows the structure and fabrication process as described by Wang et al. (2019). The experimental setup includes the microfluidic chip, connecting tubing, injection pump, and a microscope. The microfluidic injection pump used is Lange LS. PASTEURII01-1A and the microscope is Carl Zeiss Axio Observer 7. The six on-site bacterial species were individually mixed with *S. pasteurii* in a 1:1 volume ratio, resulting in a total of six mixed groups. Additionally, there was a control group where *S. pasteurii* was mixed with a blank culture medium. The initial OD$_{600}$ is 1.8±0.2 for each bacterial strain before mixing. 11.56 μL (1.25PV) of bacterial suspension is injected into the microfluidic chip at a flow rate of 0.863 μL/min. After 24 hours of settling at room temperature, 6 rounds of injection of the cementation solution are conducted at a flow rate of 0.863 μL/min, with a 24-hour interval between injections and a single injection volume of 11.56 μL.

Images of the internal structure of the chip were captured after each injection using inverted optical microscopy, and statistical analysis was conducted at the conclusion of the experiment. All microfluidic chip images were obtained using the Carl Zeiss Axio Observer 7 inverted microscope, which was equipped with a camera connected to a computer. Phase-contrast illumination and a 10× objective were utilized for imaging bacteria and calcium carbonate crystals. For the statistical analysis of bacterial quantity and crystal count, a 500 μm × 500 μm region near the chip inlet, center area, and outlet was selected, and the quantity of bacteria and crystals within each region was recorded and averaged. Crystal size testing involved the use of Zeiss Axio Vision image analysis software to analyze the images and determine the corresponding crystal area.

**Soil column experiment and MICP treatment**

To study the effects of on-site bacterial strains on the efficiency of MICP in strengthening soils, MICP was applied on a fine silica sand (purchased from Xiamen ISO Standard Sand Co.,Ltd) with an average particle size of 0.125 millimeters that adheres to the CHINA ISO standard. The sand possesses exhibits a coefficient of uniformity ($C_u$) of 5.6. According to the Unified Soil Classification System (ASTM, 2017), the sand is categorized as poorly graded. For MICP treatment one strain exhibiting the highest urease activity among the enriched bacterial cultures and another strain with the lowest low-urease activity was selected for experimentation. These strains were combined in various volumetric ratios with *S. pasteurii*. The cementation solution used was the same as described in section 2.2.

Adopting the methodology outlined by Al Qabany et al. (2012), illustrated in Figure 2, soil columns were constructed using acrylic tubes measuring 39 mm in diameter, 150 mm in height, and with a wall thickness of 2 mm. The base of these columns was sealed with a rubber stopper, which centrally incorporated a soft tube extending approximately twice the column's height. At the base of each column, a layer of porous sandstone followed by a filter paper was positioned to support the addition of sand. The sand was introduced using the dry pluviation method, with the quantity of dry sand carefully calculated to achieve a desired relative density of 60% (Tang et al. 2020) and a column fill height ranging between 78 mm and 80 mm.

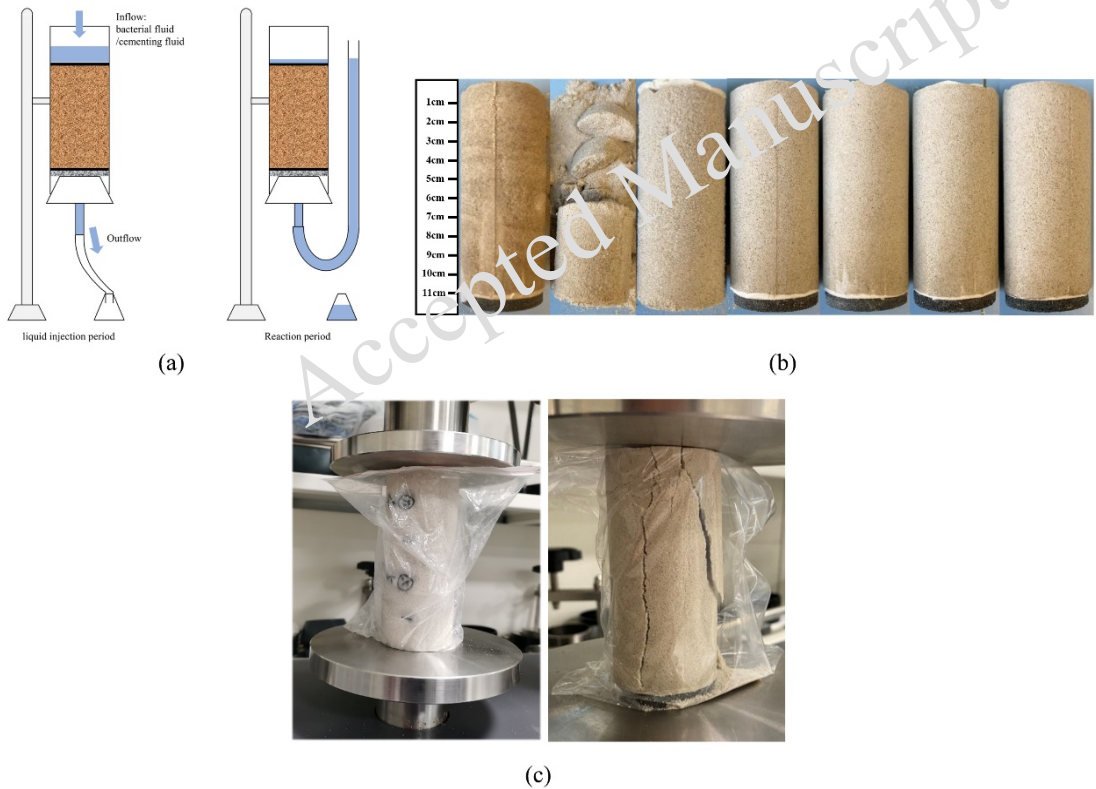

**Fig. 2.** Image of soil column experiment: (a) schematic diagram of the experimental procedure for the soil column unit; (b) soil column specimens after the experiments, from left to right: *Sp*-control, *Bs*-control, *Ph*-control, *Bs*-mix, *Ph*-mix, *Bs+Ph*-mix and in situ bacterial solution-mix; (c) unconfined compressive strength tests.

During MICP treatment, a two-phase injection process was employed. In the first phase, bacterial suspension was injected once, followed by a 6-hour retention period, while in the second phase, the cementation solution was injected six times at 24-hour intervals. Both phases utilized gravity for injection into the column. During the injection phases, the outlet tubing of the soil columns was lowered to allow liquid penetration into the soil samples via gravity. Conversely, during the injection intervals, the outlet tubing was raised to halt the flow of bacterial and cementation solution liquids. Previous research has demonstrated that gravity-assisted bacterial injection promotes the homogeneous distribution of bacterial cells (Konstantinou et al., 2023). After completing the experiment, specimens were rinsed with deionized water to remove excess ions from the pore liquid. Following MICP treatment, the sand samples were removed and placed in an oven at 105°C for at least 48 hours to dry before conducting the Unconfined Compressive Strength (UCS) test.

**Unconfined Compressive Strength (UCS)**

The unconfined compressive strength (UCS) test follows the standard testing method outlined in ASTM D2938-86 for intact rock core samples. A uniform loading rate of 1.0 mm/min was applied to the sample until the failure of the sample occurred.

**Measurement of $CaCO_3$ Content and Chemical transformation efficiency**

After UCS test, following the methodology described by Wang et al. (2019), the $CaCO_3$ content in the MICP-treated soil samples is determined using a chamber equipped with a connected pressure gauge (Wang et al., 2019b). The sand columns are divided into five sections based on their distance from the top surface. Approximately 20-30 g of soil sample from each section is ground and then placed inside the chamber. The exact mass of the ground sample is measured before placement. Subsequently, a beaker containing 30 ml of hydrochloric acid (2.5 mM) is introduced into the chamber. After sealing the chamber, agitation is employed to mix the hydrochloric acid with the soil sample, facilitating the decomposition of $CaCO_3$ into $CO_2$ and thereby increasing the chamber pressure. The mass of $CaCO_3$ within the sample is calculated using an empirical formula based on these measurements. In accordance with the research conducted by Wang et al. (2022), the definition of chemical transformation efficiency is the percentage of the actual measured mass of $CaCO_3$ relative to the theoretical calculation mass of $CaCO_3$, considering that all injected $CaCl_2$ within the soil voids is transformed into $CaCO_3$. Equation 4 is presented as follows:

$$Efficiency(\%) = \frac{m(CaCO_3)/m_1(sand)}{C(CaCl_2) \times PV \times IN \times M(CaCO_3)/m_2(sand)} \times 100\% \quad (4)$$

where *m(CaCO₃)/m₁(sand)* is the actual measured $CaCO_3$ content relative to the dry mass of sand, *C(CaCl₂)* (mol/L) is the concentration of calcium chloride in the cementitious solution, *PV* (L) is the pore volume of the sand column, *IN* is the number of injections of the cementitious solution, *M(CaCO₃)* is the molar mass of calcium carbonate, a constant value of 100 g/mol, $m_2(sand)$ (g) is the total dry mass of sand in the sand column Wang et al. (2022).

**Scanning Electron Microscope (SEM)**

To elucidate the morphological attributes and spatial disposition of precipitates, as well as to investigate the interlocking dynamics between soil particles and crystals, comprehensive microscopic analyses were conducted on the soil specimens subjected to MICP. Preceding the commencement of microscopic assessments, a minute subsample extracted from the central axis of the soil column was meticulously sectioned into block particles measuring 1 cm × 1 cm. The internal porosities of these particles were subsequently evacuated through the utilization of a vacuum drying process before undergoing scanning procedures. After the preparatory steps, microscopic examinations were conducted employing a state-of-the-art scanning electron microscope (Merlin, Zeiss).

## 3. Results and discussions

**Density and activity of the mixed microbial suspension**

On-site low-ureolysis bacteria and high-ureolysis bacteria have similar effects on bacterial growth but different effects on bacterial activity when mixed with *S. pasteurii* (see Figures 3 and 4). The control group is pure *S. pasteurii*. Three low-ureolysis bacterial experimental groups are mixtures of *S. pasteurii and three types of* low-ureolysis bacterial strains, namely *Bacillus subtilis* (*B. subtilis*), *Mesobacillus aurantius* (*M. aurantius*), and *Mammaliicoccus fleurettii* (*M. fleurettii*), respectively, at a volumetric ratio of 1:1. These three experimental groups are named *Bs*-mix, *Ma*-mix, and *Mf*-mix bellow and the relevant results are shown in Figure 3. Three urease-producing bacterial experimental groups are mixtures of *S. pasteurii and three types of* urease-producing bacterial strains, namely *Vagococcus fluvialis*, *Planococcus halotolerans* (*P. halotolerans*), and *Sporosarcina ureae* (*S. ureae*), respectively, at a volumetric ratio of 1:1. These three experimental groups are named *Vf*-mix, *Ph*-mix, and *Su*-mix bellow and the relevant results are shown in Figure 4. Experimental duration is 120 hours. Despite the slight increase in OD values of mixtures containing *S. pasteurii* with either low-ureolysis or high-ureolysis bacteria over the 120-hour period after mixing (Figure 3a and 4a), bacterial activities consistently decreased over time. This decline in activity was also observed in *S. pasteurii* without any other bacterial strains mixed in, suggesting that bacterial activity diminishes over time even when the bacterial count increase slightly. This reduction in activity may be attributed to a decrease in nutrients within the bacterial suspensions due to bacterial consumption. Comparing the bacterial activity of *S. pasteurii* (halved) with that of mixtures of *S. pasteurii* and low-ureolysis bacteria (Figure 3b), it was found that the former exhibited lower activity, whereas compared to mixtures with high-ureolysis bacteria (Figure 4b), the activity was higher. This disparity suggests that low-ureolysis bacteria inhibit *S. pasteurii*, resulting in reduced urease activity within the bacterial suspension, while urease-producing bacteria exert a stimulating effect.

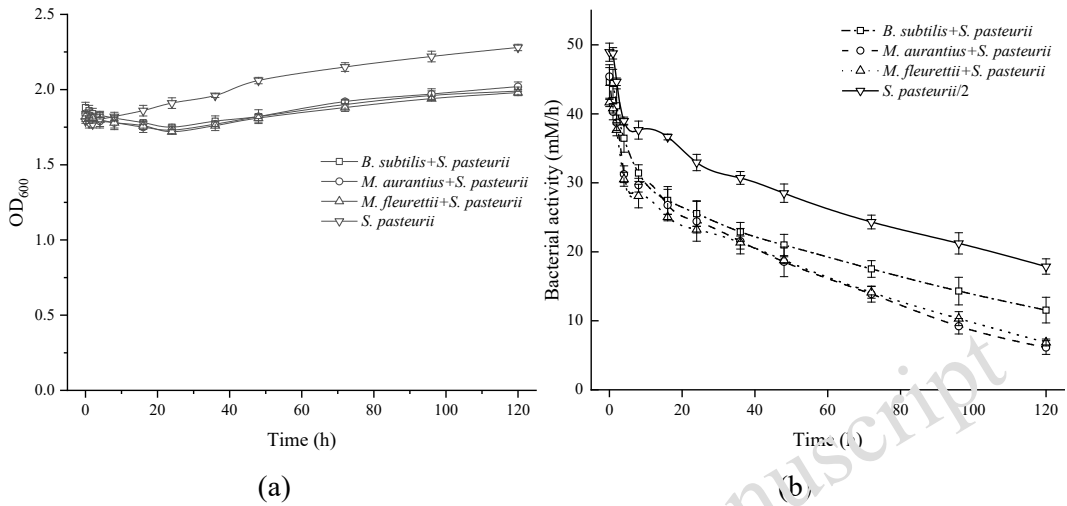

**Fig. 3.** Measurement of optical density and activity in a bacterial mixture comprising OSLUB-mix and *S. pasteurii*: (a) assessment of bacterial optical density at 600 nm ($OD_{600}$); (b) evaluation of bacterial activity.

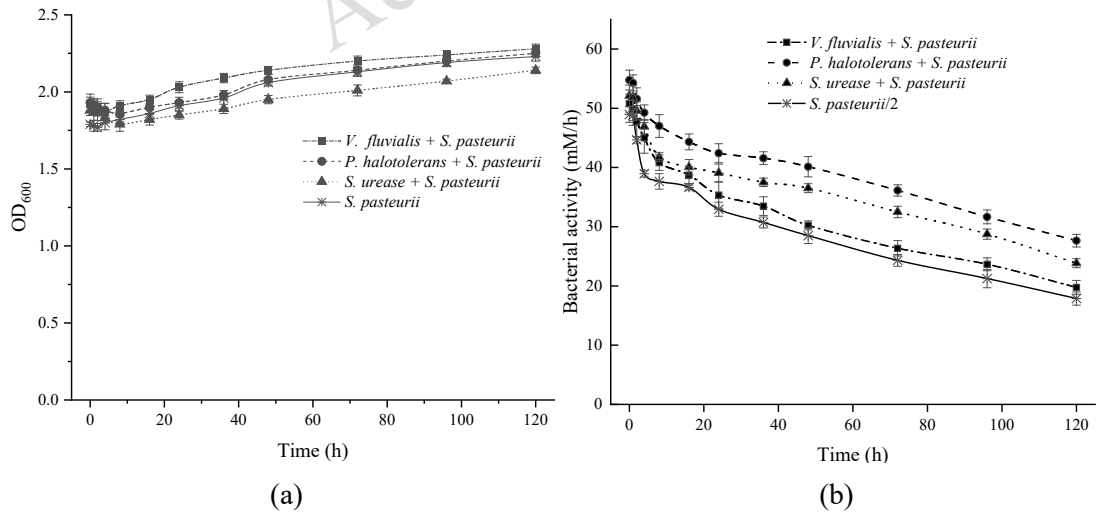

**Fig. 4.** Measurement of optical density and activity in a bacterial mixture comprising OSHUB-mix and *S. pasteurii*: (a) assessment of bacterial optical density at 600 nm ($OD_{600}$); (b) evaluation of bacterial activity.

It should be noted that given the significant variability in bacterial sizes (ranging from 8 to 20 μm, as outlined in section 3.2), relying solely on optical densities may not offer an accurate assessment of the actual bacterial count. Compounding this issue is the difficulty in distinguishing between bacterial strains in mixed bacterial cultures, making comparisons based on bacterial sizes complex and rendering optical density a somewhat imprecise indicator of bacterial abundance in the systems. Future research endeavors could explore combining liquid experiments with microfluidic techniques or employing observations of bacterial sizes and DNA analysis to quantify the abundance of each bacterial strain during growth, thus facilitating a more thorough analysis of the effects of on-site low-ureolysis and high-ureolysis bacteria on the growth of *S. pasteurii.* However, given the minimal changes observed over the 120-hour period following mixing, this study primarily focuses on analyzing the effects of mixing low-ureolysis and high-ureolysis bacteria on changes in bacterial activity. Gat et al. (2014) observed a fivefold increase in the $OD_{600}$ values of bacterial suspension, rising from 0.04 to 0.2 upon mixing *B. subtilis* and *S. pasteurii*. In contrast, this study noted only a slight increase in $OD_{600}$ values of bacterial suspension. This discrepancy may primarily stem from the significant disparity in the nutrient broth (NB) concentration utilized. Gat et al. employed a nutrient concentration of 13 g/L, markedly higher than the 3 g/L concentration used in this study. The elevated nutrient availability in their study likely contributed to the observed increase in $OD_{600}$ values following bacterial mixing. Another factor may be attributed to the growth phases of the bacteria: in Gat et al. (2014), the bacteria were still in the fast-growing phase, whereas in the present study, they were in the plateau phase. Nevertheless, despite the variations in bacterial density changes, findings from both Gat et al. (2014) and our current study suggest that low-ureolysis bacteria, such as *B. subtilis*, induce a reduction in the overall urease activity of the bacterial suspension.

In addition, Graddy et al. (2018) observed that *S. pasteurii* induces an elevation in environmental pH (alkalization) during its cultivation process. The introduction of low-ureolysis bacteria, on the other hand, due to their life processes like amino acid oxidation, results in a reduction in environmental pH, thereby inhibiting the proliferation of *S. pasteurii* (Graddy et al., 2018). These two mechanisms might explain the significant low efficiency of MICP observed in field experiments (Nasser et al., 2022). Following the pH observations of Graddy et al. (2018), the pH values of mixtures of *S. pasteurii* with either low-ureolysis bacteria or high-ureolysis bacteria were conducted and the results are shown in Figure 5 (a) and (b), respectively. It is shown that after mixing with low-ureolysis bacteria, the pH values all decrease with time, whereas after mixing with high -ureolysis bacteria the pH values all increase with time. This results are consistent with (Graddy et al. (2018) and further proved the importance of maintaining high pH for maintain high bacterial activity in bacterial suspensions.

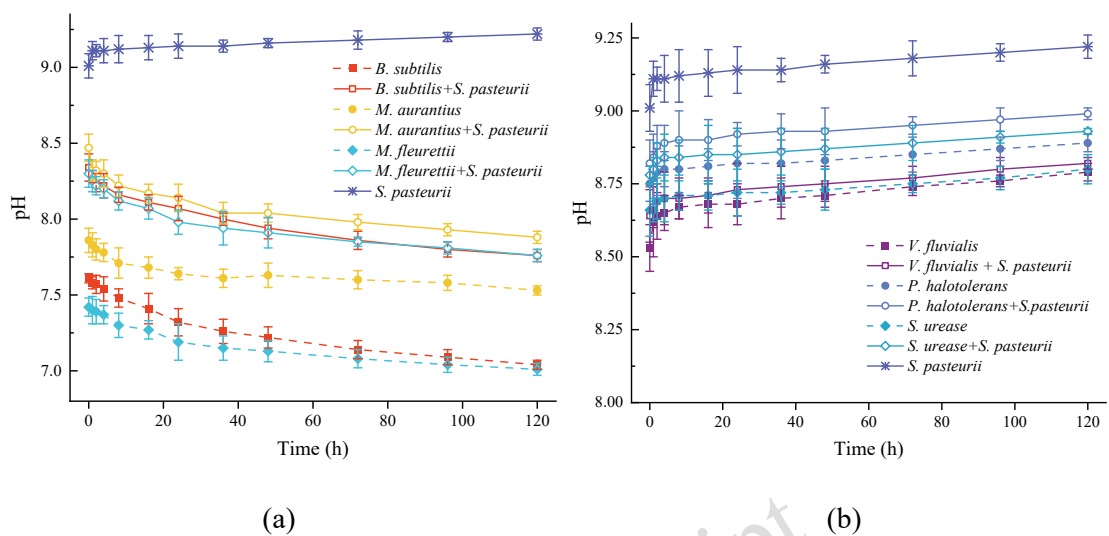

**Fig. 5.** Measuring the acidity or alkalinity (pH) of OSLUB-mix groups, OSHUB-mix groups and *S. pasteurii*: (a) OSLUB-mix groups and *S. pasteurii*; (b) OSHUB-mix groups and *S. pasteurii*.

**Bacterial growth, aggregation, and desorption**

As outlined in section 2.4, the optical density of injected bacteria for *S. pasteurii*, OSLUB-mix groups, and OSHUB-mix groups is reported as 0.9±0.1, 1.8±0.2, and 1.8±0.2, respectively. Consequently, assuming uniform bacterial sizes, the initial count of *S. pasteurii* should theoretically be half that of both the OSLUB-mix and OSHUB-mix groups (Figure 6). However, upon examination, the actual counts reveal discrepancies: the initial count of *S. pasteurii*, along with those of the OSLUB-mix and OSHUB-mix groups, are $3.8\times10^7$ cells/ml, $4.5\times10^8$ cells/ml, and $5.7\times10^8$ cells/ml, respectively. This discrepancy arises from variations in bacterial sizes, rendering the optical density an inaccurate representation of bacterial quantity. Research indicates that the volume of the three low-ureolysis bacterial cells is considerably larger (25-40 μm$^3$) than that of *S. pasteurii* cells. As a result, the mixed bacterial count is substantially lower than twice that of the *S. pasteurii* control group. Similarly, the volume of the three urease-producing bacterial cells (15-20 μm$^3$) is slightly larger than that of *S. pasteurii* cells (5-10 μm$^3$) (Yu et al., 2014; Gan et al., 2018; Rai et al., 2021), therefore, the mixed bacterial count is lower than twice that of the *S. pasteurii* control group.

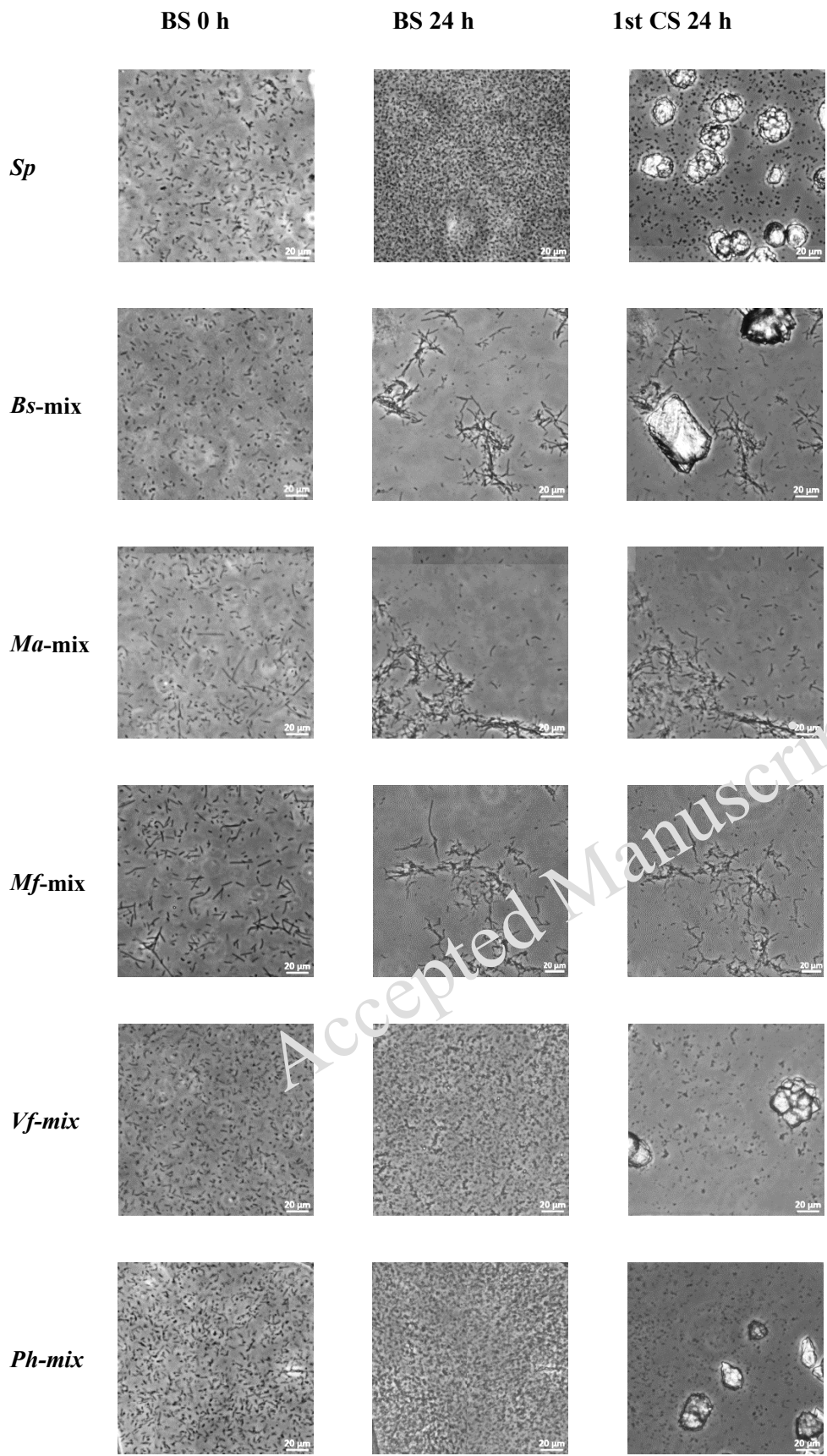

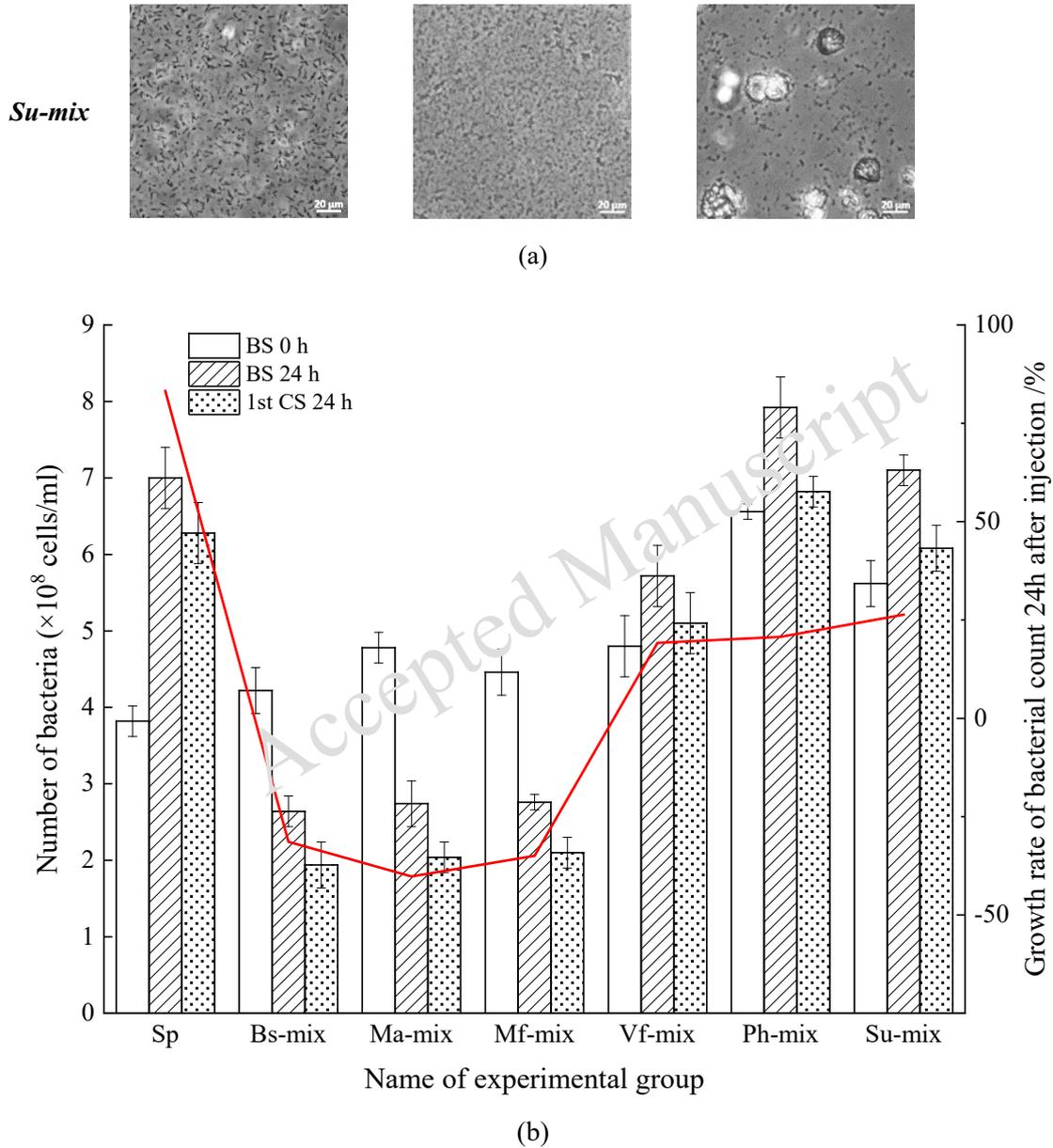

(a)

(b)

**Fig. 6.** Changes in the bacterial count for OSLUB-mix groups, OSHUB-mix groups and *S. pasteurii*: (a) sequential microfluidic chip images: post-bacterial injection, 24 hours post-bacterial injection, and 24 hours after initial cementation solution injection; (b) bacterial cell count following bacterial injection, at 24 hours post-bacterial injection, and 24 hours after initial cementation solution injection; the red line represents bacterial growth rate during the 24-hour settling period after bacterial injection.

Bacterial growth during the 24-hour setting time varied among the groups. Compared to immediately after injection, the *S. pasteurii* group exhibited an 83.4% increase in bacterial count. The OSHUB-mix groups showed increases ranging from 9.2% to 26.33% (with an average of 22.14%), while the OSLUB -mix groups experienced decreases ranging from 13.45% to 15.48% (with an average of 14.26%). This illustrated that both OSHUB-mix and OSLUB-mix have much lower growth rate than *S. pasteurii*. In contrast to Figures 3 and 4, where the initial bacterial densities are similar across all three cases and bacterial growth occurs within comparable ranges, the notable distinction in the microfluidic chip experiment arises from the growth phases of the *S. pasteurii* group versus the other two groups. While the *S. pasteurii* group remains in the growing phase, the other two groups have reached a plateau phase. Despite the challenges in directly comparing *S. pasteurii* with either OSHUB-mix or OSLUB-mix, it is evident that the OSHUB-mix group exhibits a higher growth rate, whereas the OSLUB-mix groups show a reduction in bacterial numbers. This trend is clearly depicted in Figure 6a, where during the 24-hour settling period, bacterial aggregations occur in the OSLUB-mix groups but not in the *S. pasteurii* and OSHUB-mix groups. These bacterial aggregations make it difficult to count individual bacterial cells, which could explain the reduced bacterial count observed in the OSLUB-mix groups.

Following the initial injection of the cementation solution, bacterial detachment was observed in all three cases: *S. pasteurii*, OSHUB-mix, and OSLUB-mix. However, the detachment ratios were not distinctive. Nonetheless, due to the low growth rate, the bacterial count in the OSLUB-mix group after 24 hours post-injection of cementation ($2.86 \times 10^8$ cells/ml on average) is significantly lower than either *S. pasteurii* ($6.2 \times 10^8$ cells/ml) or the OSHUB-mix groups ($6.0 \times 10^8$ cells/ml on average). The bacterial numbers continue to decrease in the subsequent injections of the cementation solution, as illustrated in Figures 7a and 7b. After the injection of the 6$^{th}$ cementation solution, the bacterial numbers for *S. pasteurii*, OSLUB-mix, and OSHUB-mix are $1.48 \times 10^8$ cells/ml, $1.0 \times 10^8$ cells/ml on average and $2.89 \times 10^8$ cells/ml on average, respectively. The decline in bacterial counts observed in all chips after the injection of the cementation solution is attributed to the flushing of loosely adhered or free-floating bacteria from the microfluidic chip due to the perfusion of the cementation solution into the chip's interior. This is consistent with the results obtained by Wang et al. (2019a). Following the results of Wang et al. (2019a), this study illustrated that bacterial strains such as OSHUB-mix, and OSLUB-mix affects the bacterial growth, aggregation and detachment, and therefore the real bacterial number during the MICP treatment procedure is difficult to predict. This results also show that after the growth and detachment procedures, the bacterial number in the OSHUB-mix is the highest among the three groups.

| 2rd CS 24 h | 3th CS 24 h | 4th CS 24 h | 5th CS 24 h |

*Sp*

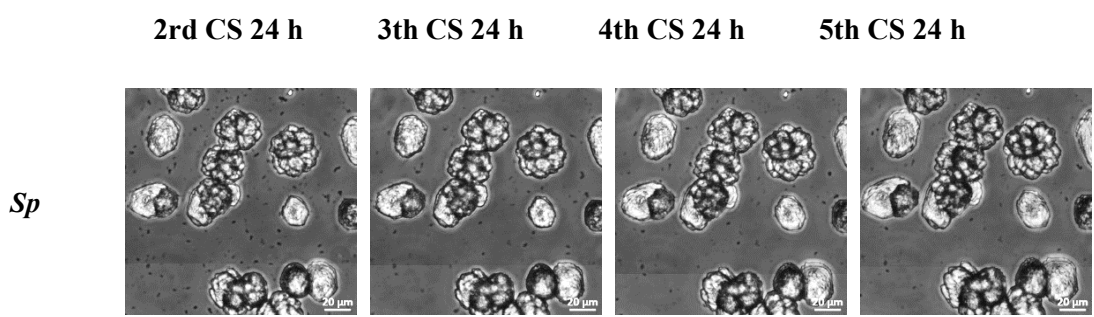

*Bs*-mix

*Ma*-mix

*Mf*-mix

*Vf*-mix

*Ph*-mix

*Su*-mix

(a)

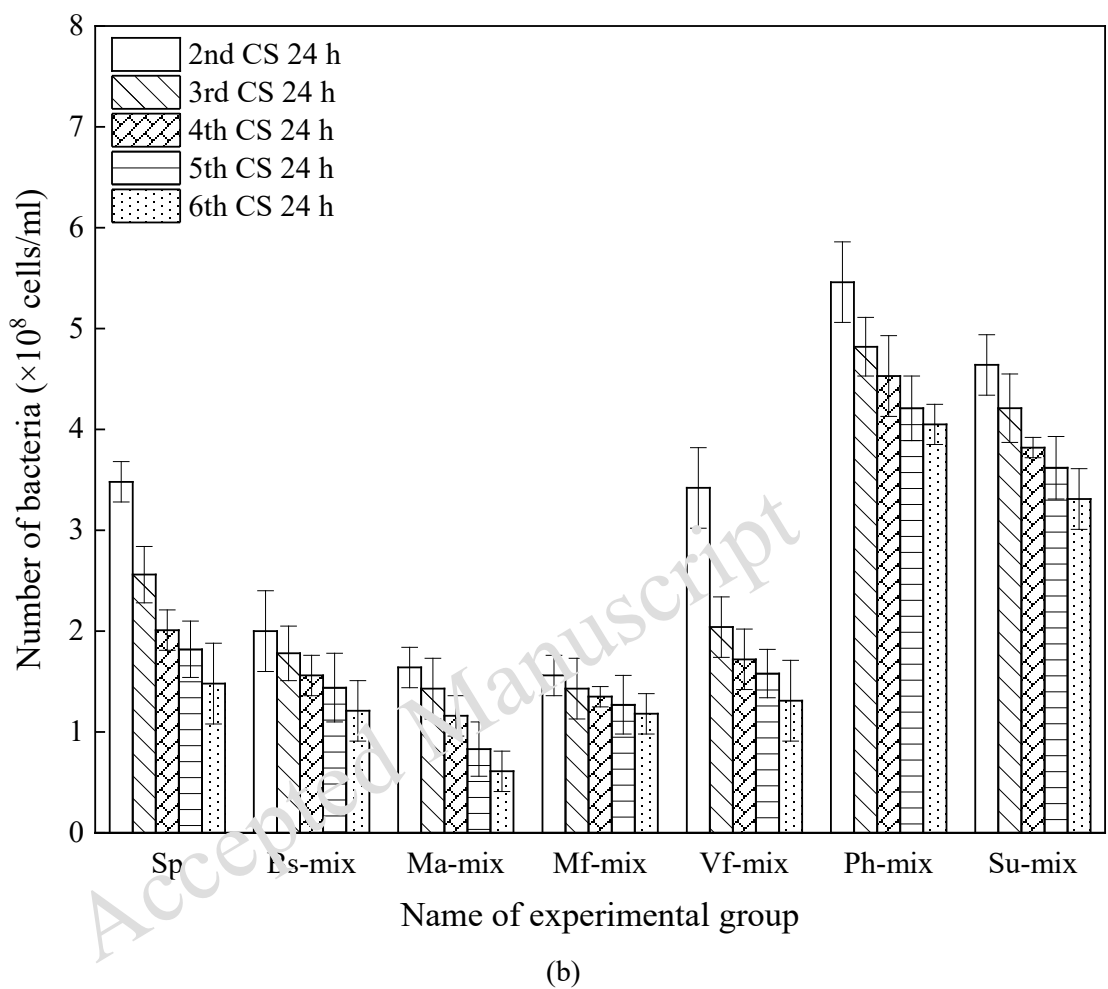

(b)

**Fig. 7.** Changes in the bacterial count for OSLUB-mix groups, OSHUB-mix groups and *S. pasteurii* (2$^{nd}$-6$^{th}$)**:** (a) microfluidic chip images captured 24 hours after subsequent cementation solution injections (200 μm×200 μm); (b) bacterial cell count following 24 hours after subsequent cementation solution injections.

**Diameter and quantity of calcium carbonate crystals**

The three OSLUB groups displayed no contribution to MICP (results not presented). Despite some ureolysis activity observed in the three OSHUB groups, their MICP performance remained severely limited (Figure 8a). Interestingly, although both OSLUB and OSHUB groups exhibited similar negligible MICP performance, their impacts on the sizes of produced calcium carbonate crystals were markedly distinct (refer to Figures 8b and 9). The diameters of calcium carbonate crystals in the three OSLUB-mix groups are in general greater than that in the control group (Figure 8 and 9a), whereas the diameters in the three OSHUB-mix groups are in general similar to control group apart from excisional case *Vf*-mix group (Figure 8 and 9b). The differences of introduced low-ureolysis bacteria or introduced high-ureolysis bacteria on crystal size are significant (Figure 9c). As the number of cementation solution injection increases, the growth rate of crystal diameter decreases rapidly (Figure 9d).

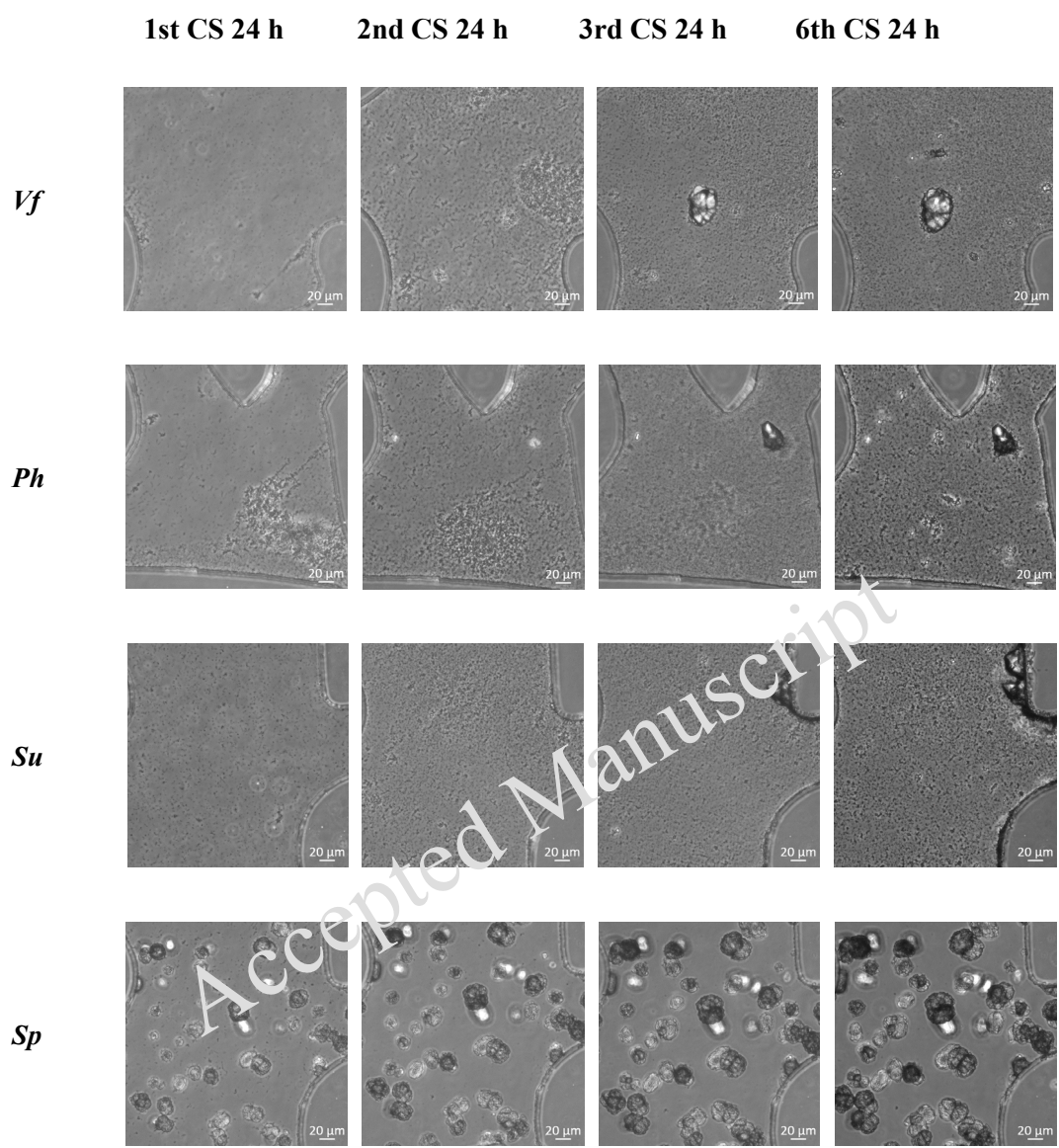

(a)

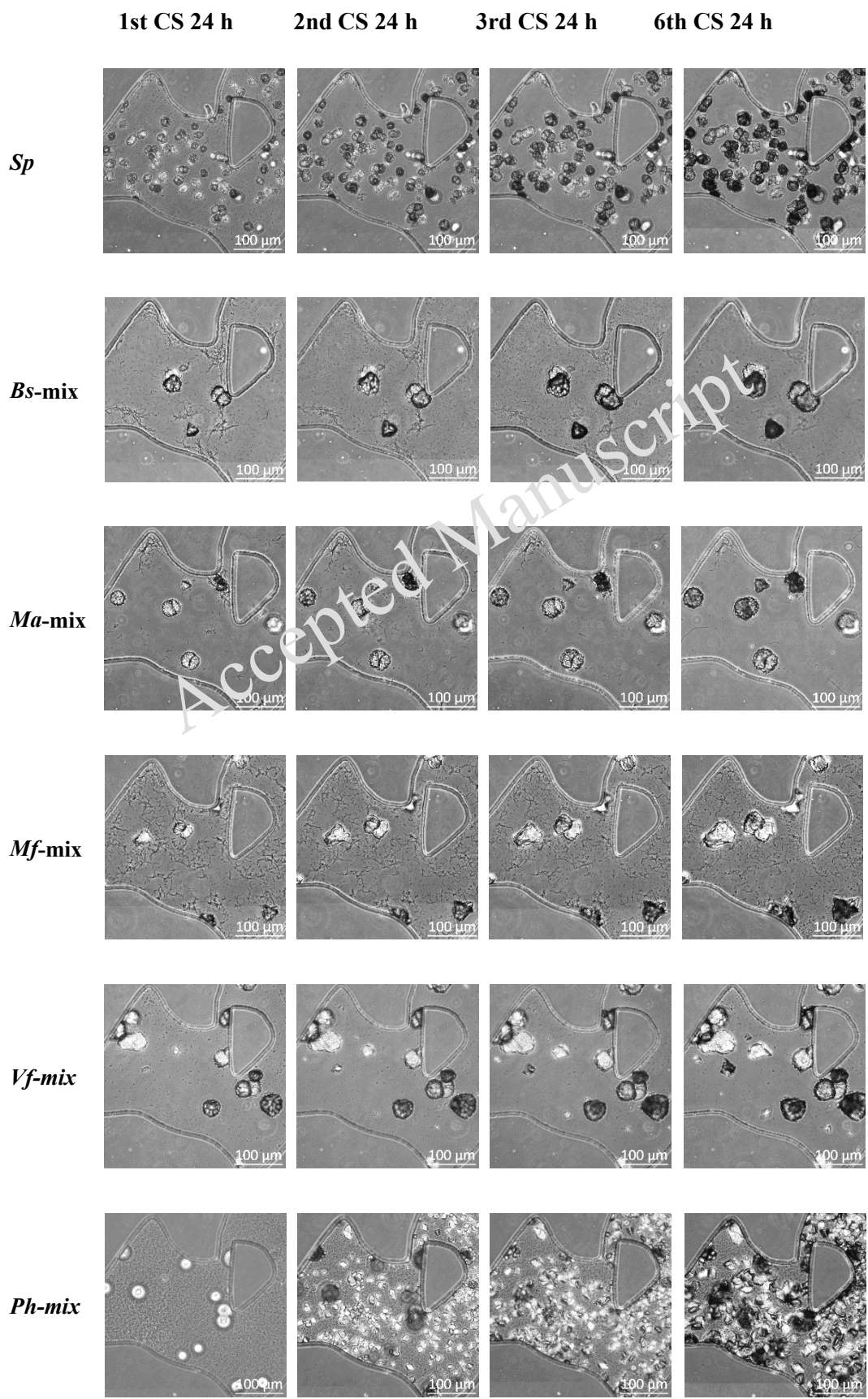

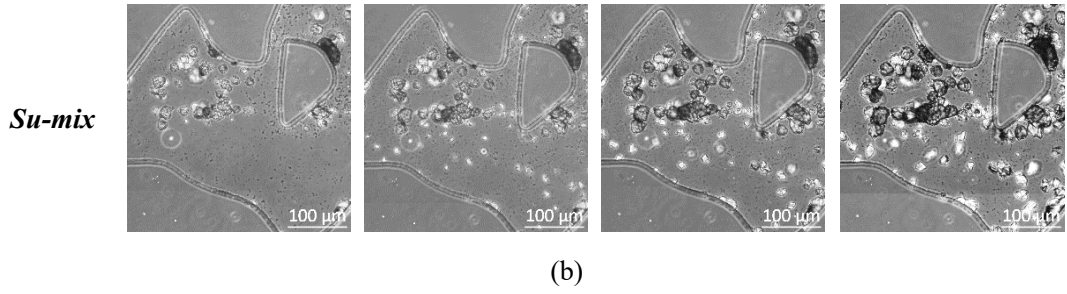

*Su-mix*

(b)

**Fig. 8.** Microfluidic chip images captured 24 hours after 1st, 2nd, 3rd, and 6th cementation solution injections (500 μm×500 μm): (a) OSHUB and *S. pasteurii*; (b) OSLUB-mix groups, OSHUB-mix groups and *S. pasteurii*

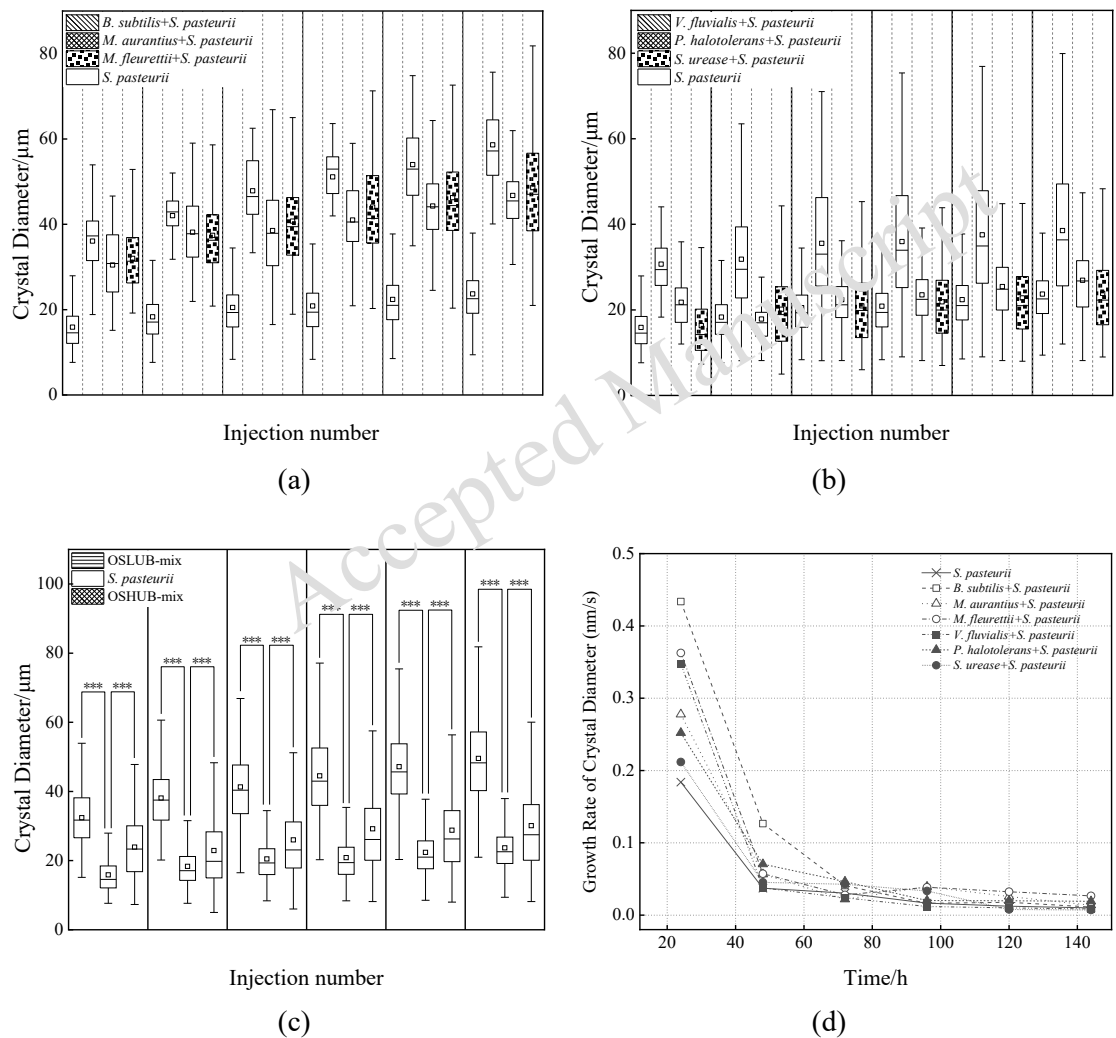

(a)

(b)

(c)

(d)

**Fig. 9.** Alterations in calcium carbonate crystal diameter produced by OSLUB-mix groups, OSHUB-mix groups and *S. pasteurii*: (a) crystal diameter of *S. pasteurii* and OSLUB-mix groups; (b) crystal diameter of *S. pasteurii* and OSHUB-mix groups; (c) comparison of crystal diameter of OSLUB-mix, OSHUB-mix and *S. pasteurii*; (d) temporal fluctuations in the growth

rate of crystal diameter.

OSLUB and OSHUB also significantly affect the numbers of produced calcium carbonate crystals (see Figures 10 and 11). The crystal number in the three OSLUB-mix groups is only about 10% of that in the control group (Figures 10 and 11a), whereas apart from one OSHUB-mix group, the crystal number in the other two OSHUB-mix group are similar to that of *S. pasteurii* group(Figures 10 and 11b). The effects of OSHUB on crystal number are less significant than OSLUB (Figure 11c). The growth rate of crystals number also displays a rapid initial decline as the injection of cementation solution increases, followed by stabilization (Figure 11d).

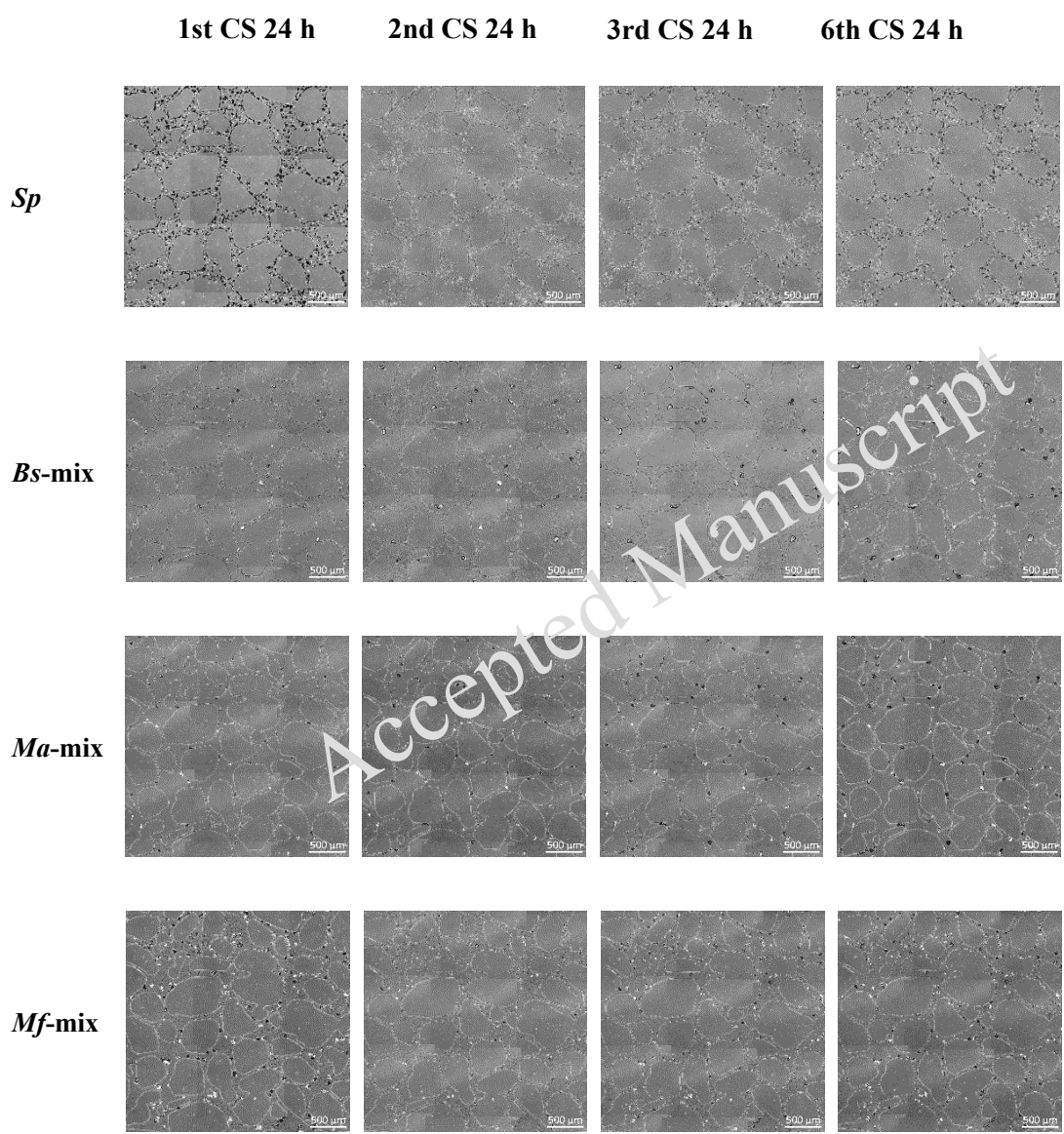

|  | **1st CS 24 h** | **2nd CS 24 h** | **3rd CS 24 h** | **6th CS 24 h** |

*Sp*

*Bs*-mix

*Ma*-mix

*Mf*-mix

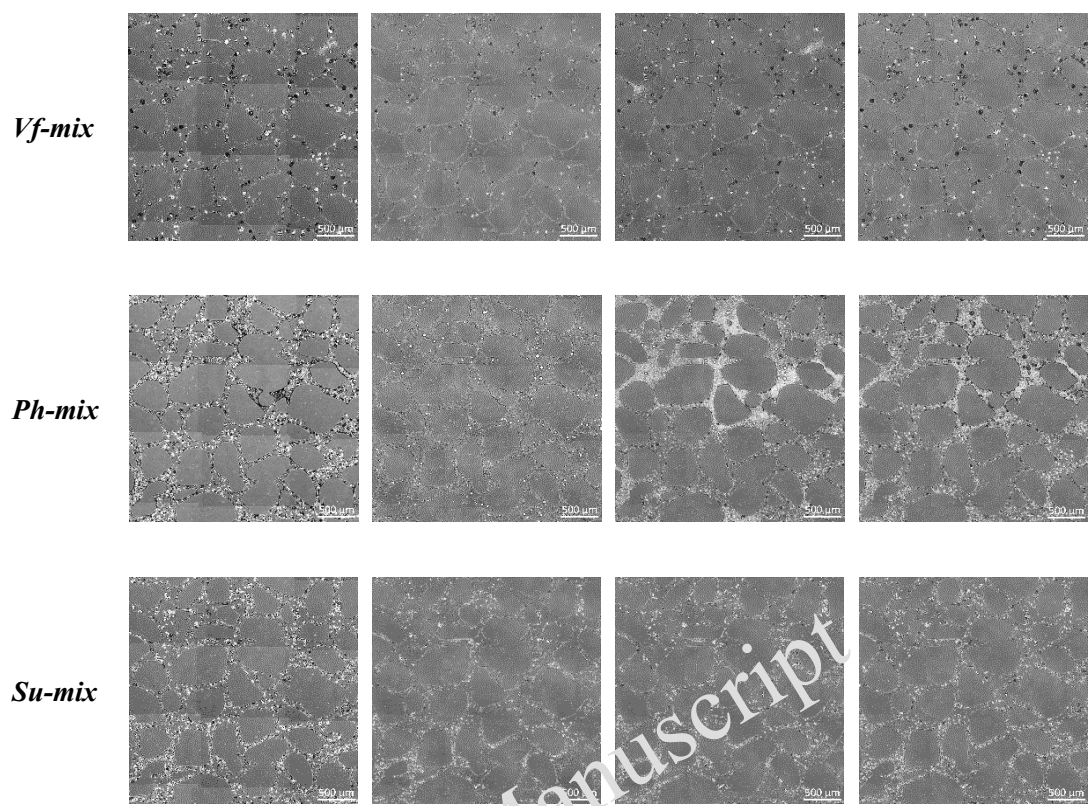

**Fig. 10.** Microfluidic chip images captured 24 hours after 1st, 2nd, 3rd, and 6th cementation solution injections (3000 μm×3000 μm).

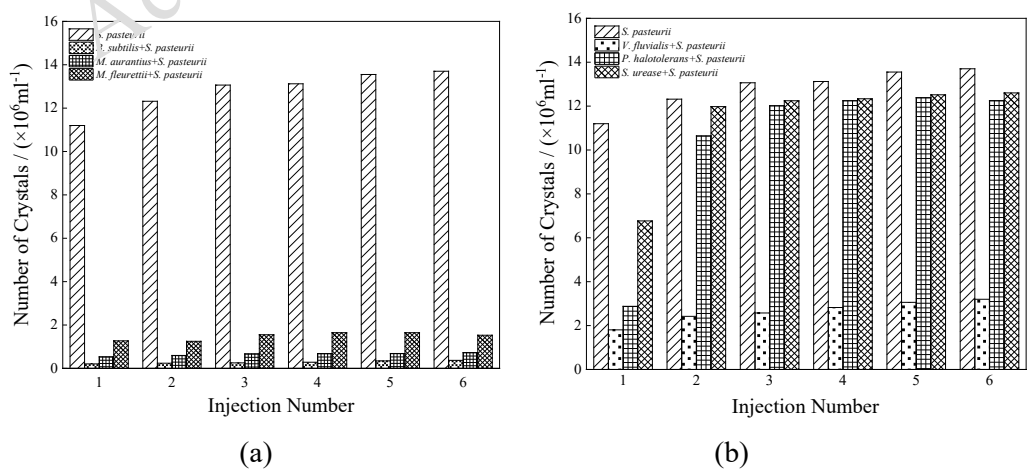

(a)                                   (b)

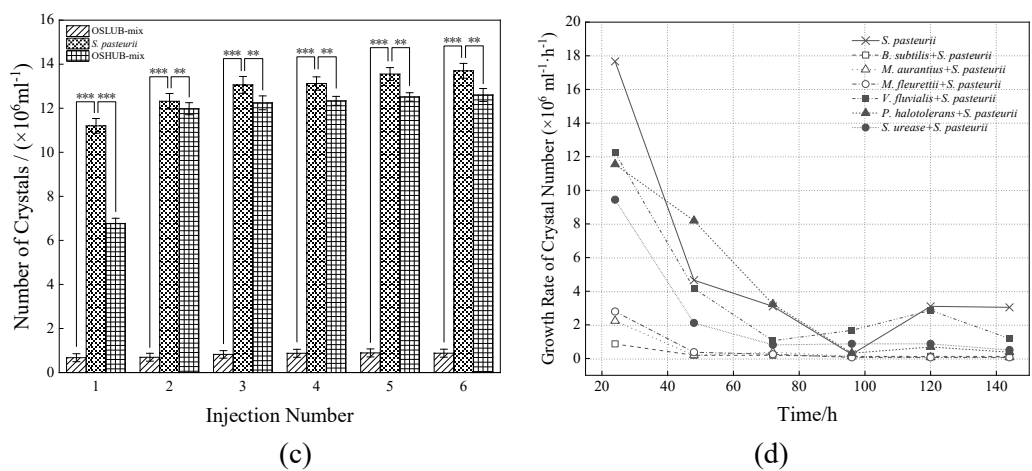

**Fig. 11.** Alterations in calcium carbonate crystal number produced by OSLUB-mix groups, OSHUB-mix groups and *S. pasteurii*: (a) crystal number of *S. pasteurii* and OSLUB-mix groups; (b) crystal number of *S. pasteurii* and OSHUB-mix groups; (c) comparism of crystal number of OSLUB-mix, OSHUB-mix and *S. pasteurii*; (d) temporal fluctuations in the growth rate of crystal number.

Previous research has shown that both bacterial density (Wang et al., 2021) and bacterial activity (Konstantinou et al., 2021) affect crystal size. When bacterial densities are low, the overall activity is reduced, causing the mineralization system to favor the production of fewer but larger (> 8000 μm$^3$) calcium carbonate crystals. Conversely, higher bacterial concentrations result in increased overall activity, leading to the generation of a larger number of smaller-sized (< 2000 μm$^3$) crystals. In this experiment, the initial urease activity of the three OSLUB-mix was 9.32 mM/h, 7.32 mM/h, and 4.63 mM/h, respectively, all were lower than the activity of *S. pasteurii* at 18.7 mM/h. The introduction of low-ureolysis bacteria reduce the overall hydrolysis activity of the system. As a result, the three OSLUB-mix groups exhibit a preference for producing calcium carbonate crystals with larger diameters in the range of 40-50 μm.

Among the three OSHUB-mix groups, the Vf-mix group formed relatively larger crystals than the other two. This phenomenon could be attributed not to a decrease in bacterial activity (as shown in Figure 4b) but to a reduced number of bacteria (illustrated in Figure 7). Specifically, Figure 7 highlights a sharp reduction in the population of the high-ureolysis bacteria within the *Vf-mix* group following the second injection of the cementing solution, with bacterial numbers falling to less than half compared to the other two high-ureolysis bacteria mix groups during the same timeframe. Meanwhile, Figure 4b indicates that the activity levels across the three OSHUB-mix groups were relatively similar, with the *Ph-mix* group, which exhibited the highest bacterial activity 48 hours after mixing, being only 22.3% more active than the *Vf-mix* group, which had the lowest activity. These findings suggest that a decrease in the number of bacteria impacts the overall urease activity of the group more significantly than does the variation in urease activity among the different bacterial strains.

**Chemical transformation efficiency of calcium carbonate**

The ratios of the volumetric fraction of calcium carbonate crystals to pore volume (Vc/Vv) and the chemical transformation efficiency (CTE) both serve as indicators of the total calcium carbonate content. For all three OSLUB-mix groups, these values are lower than those of the S. pasteurii group. In contrast, for all three OSHUB-mix groups, both Vc/Vv and CTE values surpass those of the S. pasteurii group, as shown in Figure 12. Initially, the Vc/Vv and CTE values for the OSHUB-mix groups are on par with those of the S. pasteurii group. However, after 2-3 injections of the cementation solution, the Vc/Vv and CTE for the OSHUB-mix groups exceed those of the S. pasteurii group, as detailed in Figure 12. Conversely, the gap in Vc/Vv and CTE between the OSLUB-mix groups and the S. pasteurii group widens with each injection. This suggests that mixing S. pasteurii with high-ureolysis bacteria significantly enhances and stabilizes its urea hydrolysis capability, whereas the presence of low-ureolysis bacteria increasingly inhibits the urease activity of S. pasteurii with more cementation solution injection cycles. Hence, for practical MICP applications on site soils, it is advisable to analyze the indigenous bacterial strains beforehand. Additionally, implementing methods to mitigate the presence of low-ureolysis bacteria prior to MICP treatment becomes imperative.

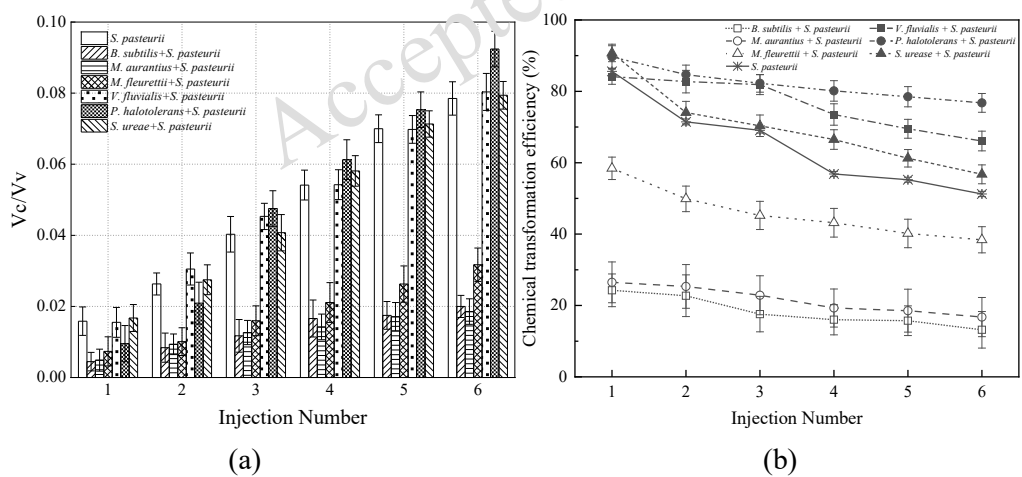

**Fig. 12.** MICP reaction efficiency: (a) ratio of crystal volume to microfluidic chip pore volume; (b) variation in the chemical conversion rate of calcium carbonate with the number of injections.

## 3.5 Morphological characteristics of calcium carbonate crystals

The morphological characteristics and distribution patterns of induced calcium carbonate crystals by control and low-ureolysis bacteria were notably distinct from the groups mixed with high-ureolysis bacteria. To elucidate these variations, the low-ureolysis bacteria mix group represented by the lowest conversion efficiency, i.e., *Bs*-mix, and the high-ureolysis bacteria mix group characterized by the highest conversion efficiency, i.e., *Ph*-mix, were selected for comparative analysis alongside the control group (Figure 13). Small crystals primarily exhibited bidirectional growth tendencies with occasional instances of tridirectionally growth (indicated by the orange and green color squire in Figure 13) produced in the *Bs*-mix group. Conversely, the *Ph*-mix group and control group generated three types of crystals: (1) small crystals which were unstable and dissolved 24 hours after cementation solution injection (indicated by red color squire in Figure 13); (2) cluster-like crystals which exhibited consistent growth (indicated by blue color squire in Figure 13); (3) dark-colored irregularly spherical crystals which exhibited a volumetric increase in a radially expanding manner with each cementation solution injection (indicated by yellow color squire in Figure 13).

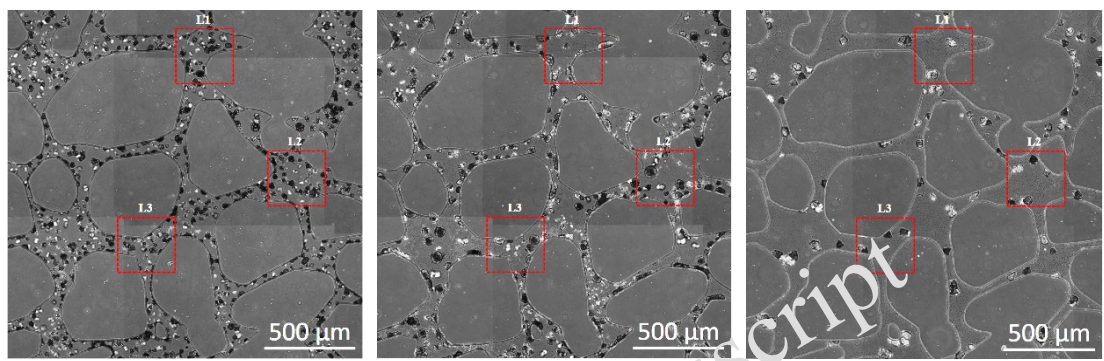

*S. pasteurii*        *P halotolerans + S. pasteurii*        *B. subtilis + S. pasteurii*

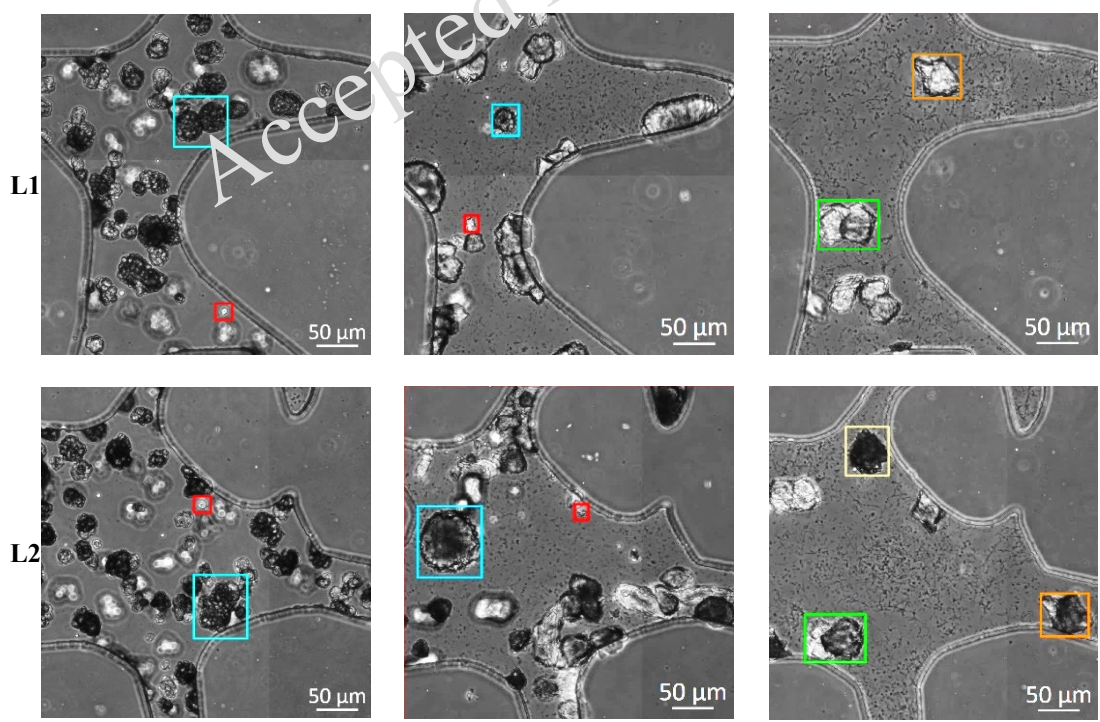

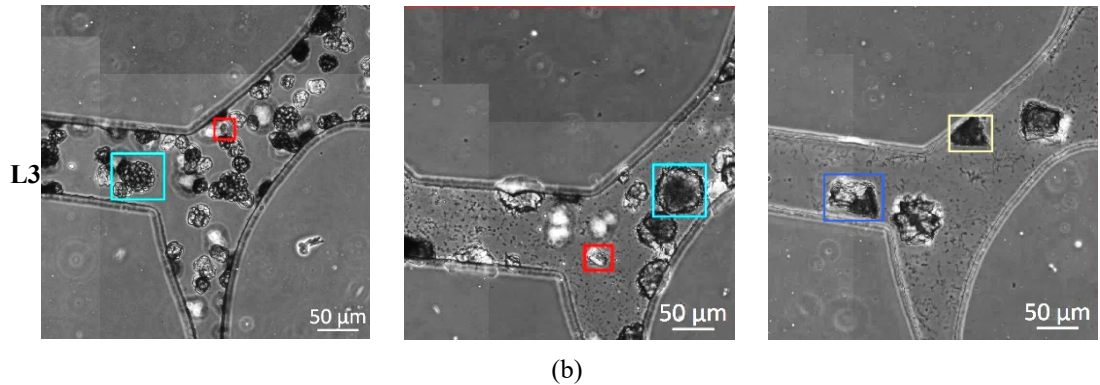

(b)

**Fig. 13.** Microfluidic chip crystal layout after 6[th] injection: (a) crystal distribution in OSLUB-mix group (*B. subtilis* + *S. pasteurii*), OSHUB-mix group (*P. halotolerans* + *S. pasteurii*) and *S. pasteurii*; (b) crystal morphology in these groups.

Prior studies such as those by Wang et al. (2021) and Konstantinou et al. (2021) have suggested that the activity of the bacterial suspension used in MICP treatment may impact the size and distribution of crystals, thereby influencing the mechanical properties of MICP-treated samples. However, as of now, no research has specifically investigated the effects of bacterial activity or other factors on crystal morphology. Further investigation could involve observing nucleation and crystal growth within porous media to elucidate the spatial distribution mechanisms of calcium carbonate at the particle scale.

**UCS of MICP-treated soils**

After MICP treatment, except for the low-ureolysis bacteria *Bs* group, no fractures were observed in the remaining specimens (Figure 2). All other specimens displayed a characteristic tensile failure pattern along the longitudinal axis (Figure 2), consistent with previous observations (Al Qabany et al., 2012; van Paassen et al., 2010; Wang et al., 2019b). Among the single bacteria, *S. pasteurii*, *B. subtilis* and *P. halotolerans*, *S. asteurii* shows the best MICP performance, followed with *P. halotolerans*, whereas *B. subtilis* almost had no MICP efficiency (Figure 14 a). Apart from the *Ph-mix* group, all the other mix groups reduced UCS values and CTE (data of the right four columns compared to the left first column in Figure 14 a and b). High-ureolysis bacteria enhance MICP performance whereas low-ureolysis bacteria inhibit MICP performance. In general, the effects of low-ureolysis bacteria on inhibiting MICP performance are more significant that the enhancing effects of high-ureolysis bacteria on MICP performance, therefore the performance of mixture of *S. pasteurii* with both low-ureolysis bacteria and high-ureolysis is about half of that of *S. pasteurii* (Figure 14c). UCS in general increase with ureolytic activity (Figure 14a), which explains why in experimental samples with higher content of non-ureolytic bacteria, the strength of the specimens is observed to be lower.

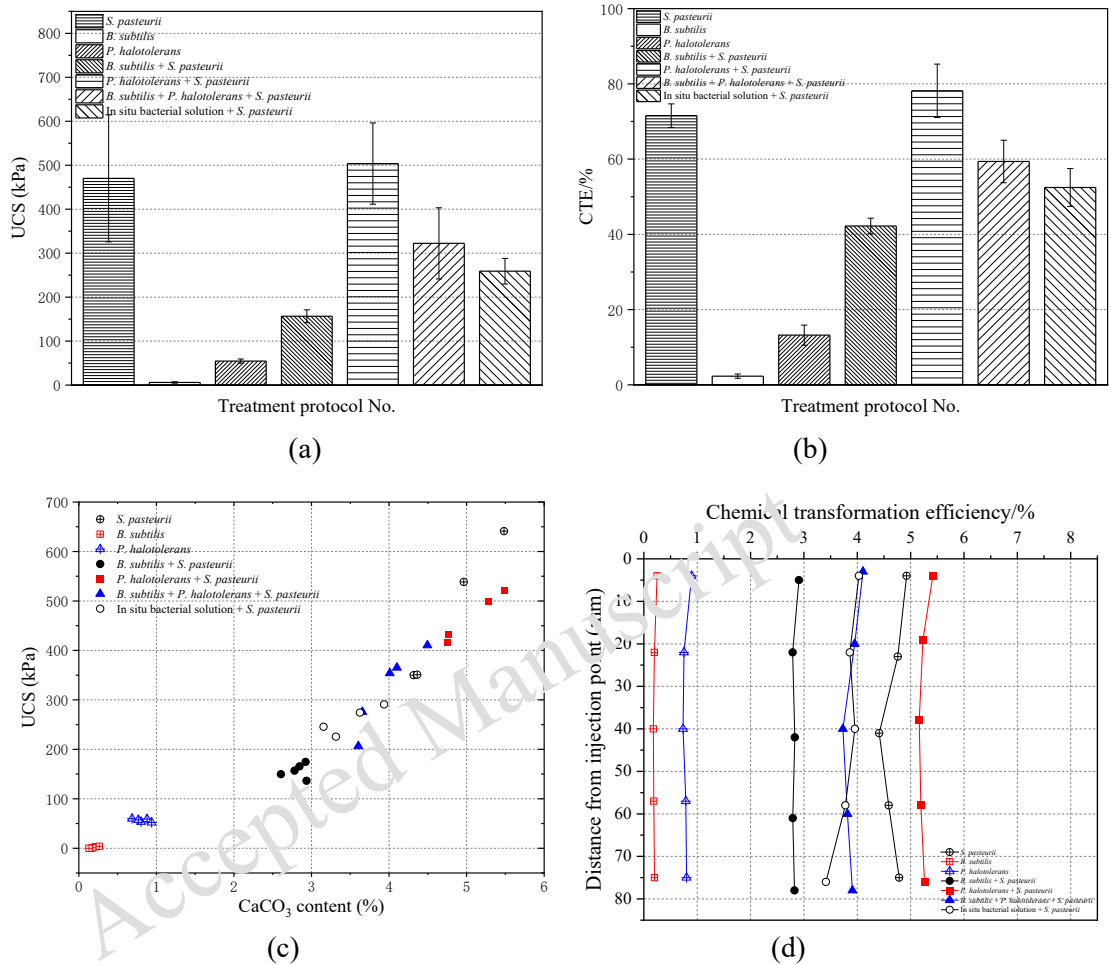

**Fig. 14.** Key physical parameters of sand column samples following micp treatment:

(a) UCS measurements of sand columns subjected to MICP treatment; (b) chemical transformation efficiency of sand samples treated with MICP; (c) correlation analysis between calcium carbonate content and UCS; (d) distribution of calcium carbonate content along an 80 mm sand column following MICP treatment.

Following the UCS test, soil samples were uniformly divided into five sections based on their distance from the top. The calcium carbonate content was measured for each section (Figure 14d). *Ph*-mix group, exhibited the most effective reinforcement. Graddy et al. (2018) observed that the introduction of *S. pasteurii* into *in-situ* soil samples resulted in the extremely low content of *S. pasteurii* in the collected effluent after one experimental cycle. The on-site bacteria within the soil potentially exerted interference and inhibition upon the proliferation of artificially introduced *S. pasteurii*. In some cases, the on-site bacteria might have even caused the demise of *S. pasteurii*, consequently leading to an overall reduction in urease enzyme activity within the bacterial solution. It was also found by Graddy et al. (2018) that the top five most abundant bacterial species in the effluent were low-ureolysis bacteria. In the present experiment, the in situ bacterial solution- +*S. pasteurii* mixed group and the low-ureolysis bacteria *Bs*-mix group exhibit comparable overall chemical conversion rates (Figure 14b), it can be speculated that following the introduction of nutrients into the soil, a substantial portion of these nutrients were preferentially consumed by low-ureolysis bacteria.

**Scanning Electron Microscope (SEM) Images**

The crystal diameter in the *Sp*-control group is generally smaller by about 20 μm (Figure 15a), which is consistent with the findings of Al Qabany et al (2012). The low-ureolysis bacteria *Bs*-control group exhibits minimal calcium carbonate crystal formation (Figure 15b), while the high-ureolysis bacteria *Ph*-control group shows a scant amount of calcium carbonate crystals adhering to the surface of soil particles (Figure 15c). In the low-ureolysis bacteria *Bs*-mix group, a small quantity of larger diameter crystals (30-40 μm) forms and adheres to the surface of soil particles, with fewer crystals at particle junctions (Figure 15d). The crystal sizes in the high-ureolysis bacteria *Ph*-mix group are comparable to those in the *Sp*-control group. However, the bonding between soil particles appears slightly inferior to that of the *Sp*-control group (Figure 15e). The crystal diameter generated in the mixed group of low-ureolysis bacteria and high-ureolysis bacteria *Bs+Ph*-mix group is notably larger than that of the *Sp*-control group, yet slightly smaller than the low-ureolysis bacteria *Bs*-mix group (Figure 15f). The calcium carbonate crystals formed in the in situ bacterial solution mixed group exhibit a wider size range (20-40 μm) and a slightly lower quantity compared to the high-ureolysis bacteria *Ph*-mix group (Figure 15g). As previously mentioned, the introduction of exogenous low-ureolysis bacteria into *S. pasteurii* can influence its MICP performance efficiency, resulting in the formation of larger but fewer CaCO$_3$ crystals.

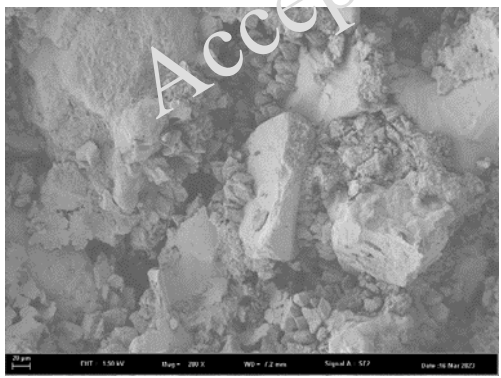
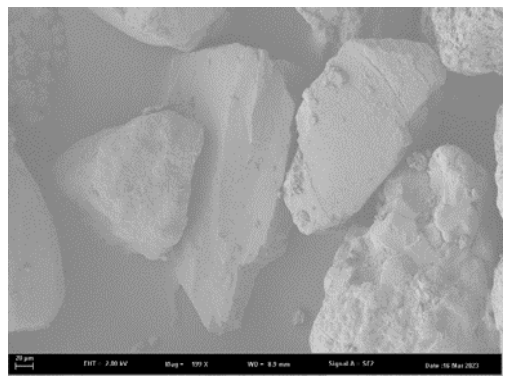

(a) *S. pasteurii*　　　　　　　　　　　　　(b) *B. subtilis*

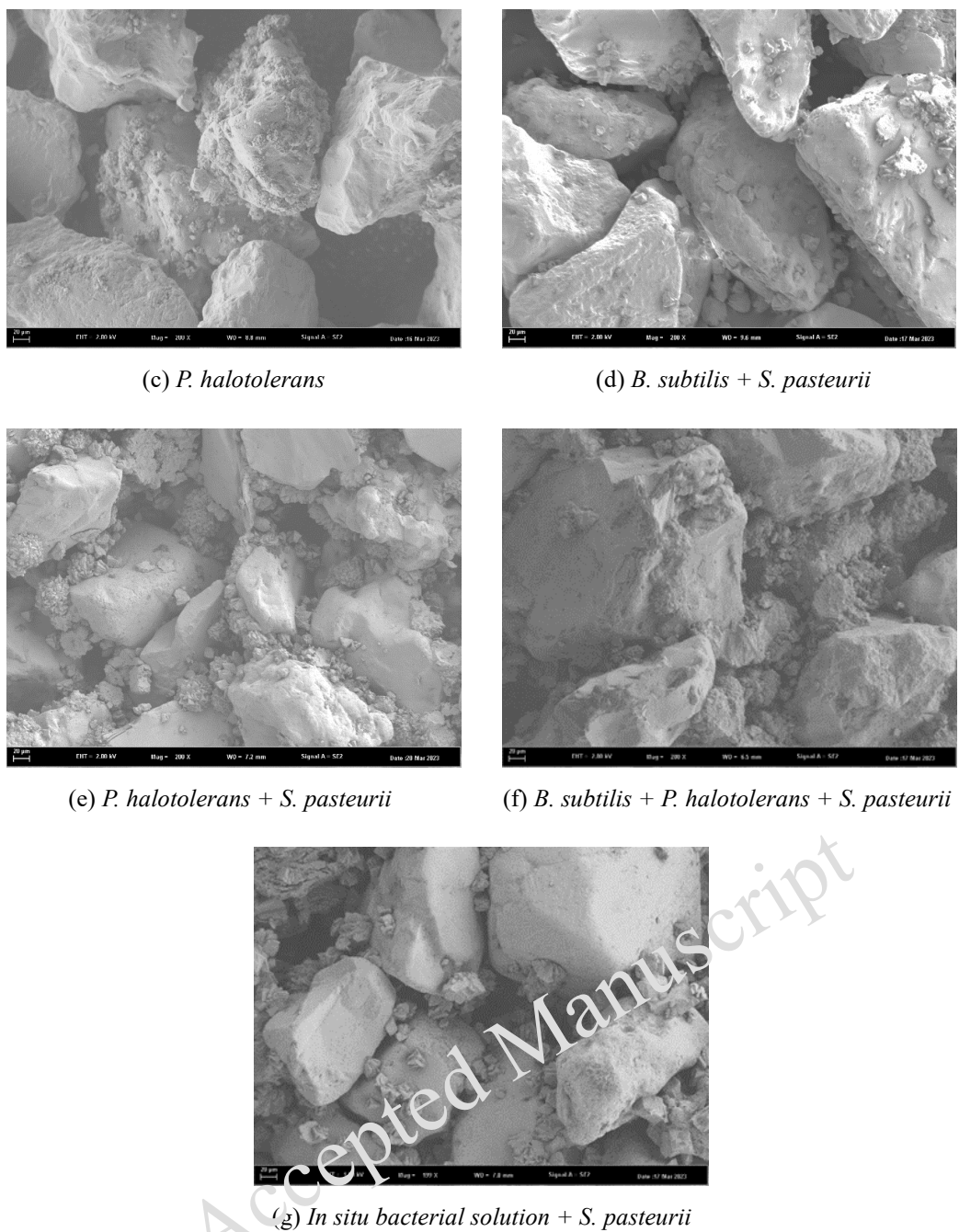

(c) *P. halotolerans*  (d) *B. subtilis + S. pasteurii*

(e) *P. halotolerans + S. pasteurii*  (f) *B. subtilis + P. halotolerans + S. pasteurii*

(g) *In situ bacterial solution + S. pasteurii*

**Fig. 15.** Scanning electron microscope (SEM) imagery of sand column samples, labeled as a-g, corresponding sequentially to the control group and the six mixed groups.

SEM images of the samples are constant with the images microfluidic chip experiment (Figure 16), indicating microfluidic chip experiment is a useful tool to study not only the processes, affecting factors of MICP (Wang et al., 2019a; Wang et al., 2019b; Wang et al., 2023), but also the morphology and characteristics of $CaCO_3$ crystals. Crystal morphologies of the *Sp*-control group and the *Ph*-mix group are both rhombus-shaped with similar sizes. However, the low-ureolysis bacteria *Bs*-mix group produces a small quantity of cluster-like, large rhombus-shaped crystals with diameters exceeding 30 μm. This is consistent with the findings of Wang et al. (2019) that lower bacterial activity tends to form lower amounts of larger $CaCO_3$ crystals.

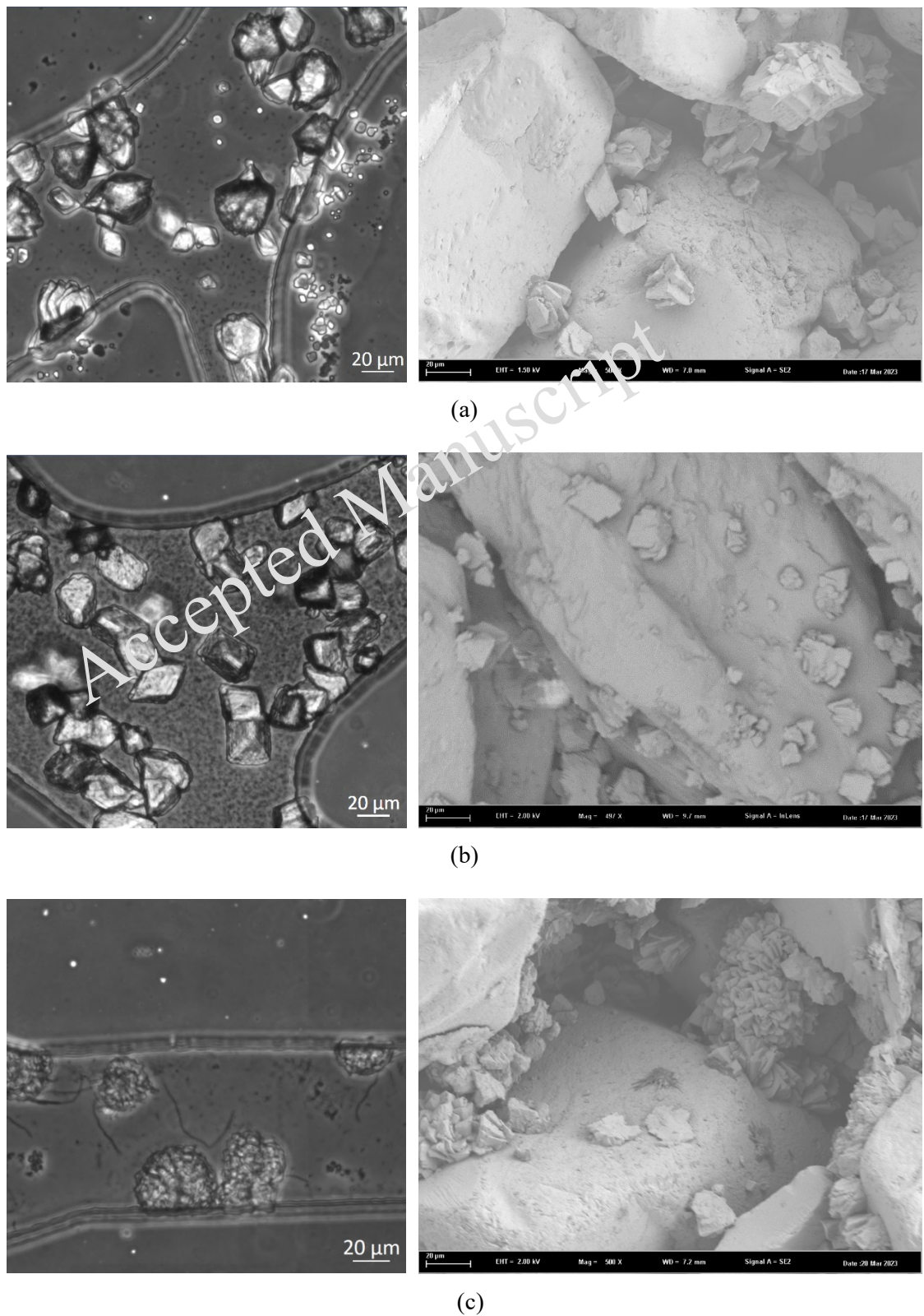

(a)

(b)

(c)

**Fig. 16.** Contrastive examination of microfluidic chip images and scanning electron microscopy (SEM) images. On the left, the image showcases the interior of the microfluidic chip experiment,

while the right side exhibits an SEM depiction of the sand column sample: (a) comparison of images for *S. pasteurii*; (b) comparison of images for OSHUB-mix group (*P. halotolerans* + *S. pasteurii*); (c) comparison of images for OSLUB-mix group (*B. subtilis* + *S. pasteurii*).

# 4. Conclusions

In this study, *in-situ* bacterial isolation, 16s rDNA detection, cultivation and activity test, microfluidic chip experiments, sand column experiments, and SEM imaging were conducted to study the different effects of *S. pasteurii*, on-site low-ureolysis bacteria (OSLUB) and on-site high-ureolysis bacteria (OSHUB) on MICP efficiency. The main findings are summarized as bellow:

Bacterial mixtures liquid experiment show that comparing the three bacterial systems, namely, *S. pasteurii*, OSLUB-mix, and OSHUB-mix, despite the activity all reduced with time in the three systems due to the consummation of nutrient, the activity reduced to the lowest level after 120 hours in the OSLUB-mix, whereas reduced to the highest level in the OSHUB-mix system. This means the OSLUB decrease the activity of mixing system, whereas OSHUB helps to increase the activity of mixing system. pH experiments illustrates that in the OSLUB-mix system the pH level reduces with time whereas in the OSHUB-mix system the pH increase with time. In addition, –microfluidic experiments indicate that the bacterial quantity of OSLUB-mix groups is significantly lower than that of OSHUB-mix and *S. pasteurii* control group.

Experiments using both microfluidic chips and soil columns demonstrated that groups treated with OSHUB-mix exhibited significantly higher MICP chemical transformation efficiency and soil strength enhancement performance than those treated with OSLUB-mix, and even outperformed the *S. pasteurii* group slightly. This suggests that although bacteria with high ureolytic activity alone may not achieve MICP efficiency on par with *S. pasteurii*, their presence in the system does not hinder the performance of *S. pasteurii*. Conversely, bacteria with low ureolytic activity not only fail to contribute to MICP efficiency but also notably impede the performance of *S. pasteurii*.

The microscopic images on microfluidic chip samples and SEM images on soil column samples are consistent and both show that the crystals in the OSLUB-mix groups are larger in size and fewer in number, and the chemical conversion rate of calcium carbonate is lower than OSHUB-mix groups. This result is consistent with previous published results which indicate that lower bacterial activity produces larger but fewer carbonate crystals. While OSHUB-mix groups tend to produce larger crystals, potentially enhancing soil particle bonding, the significantly lower total content of carbonate crystals precipitated results in a much weaker overall strength enhancement performance compared to OSLUB-mix groups.

This study elucidates that the rapid decline in MICP efficiency of *S. pasteurii* after introducing it into the on-situ soil is attributed to the interference from in-situ low-urease bacteria in various aspects such as bacterial proliferation, bacterial activity, mineralization capacity, etc. It was further discovered that low-urease bacteria affect the behavior of *S. pasteurii* by altering the environmental pH. This provides a fundamental study for reducing the interference of in situ bacteria, stabilizing the environmental parameters in the operation area and improving the efficiency of MICP field experiments in the future by means of soil pre-treatment to decrease the proportion of low-urease bacteria or pH control of the pore fluid of treated soils.